\documentstyle[prd,preprint,aps]{revtex}
\tighten

\def\be{\begin{eqnarray}}
\def\en{\end{eqnarray}}
\def\bea{\begin{eqnarray}}
\def\ena{\end{eqnarray}}

\def\sg{\sqrt{-g}}
\def\sgm{\sqrt{-\gamma}}

\begin{document}
\title{THE ENERGY-MOMENTUM TENSOR FOR THE GRAVITATIONAL FIELD.}
\author{S.V.~Babak$^{1)}$ and L.P.~Grishchuk $^{1)2)}$}
\address{ 1) Department of Physics and Astronomy, Cardiff University,
Cardiff CF24 3YB, UK.}
\address{ 2)  Sternberg Astronomical Institute, Moscow University,
Moscow 119899, Russia \\
{\rm E-mail: babak@astro.cf.ac.uk ;  grishchuk@astro.cf.ac.uk}}
\maketitle
\begin{abstract}
	The search for the gravitational energy-momentum tensor is often 
qualified as an attempt of looking for ``the right answer to the wrong
question''. This position does not seem convincing to us. We think that
we have found the right answer to the properly formulated question. We have 
further developed the field theoretical formulation of the general relativity
which treats gravity as a non-linear tensor field in flat space-time. The
Minkowski metric is a reflection of experimental facts, not a possible 
choice of the artificial ``prior geometry''. In this approach, we have
arrived at the gravitational energy-momentum tensor which is: 1) derivable 
from the Lagrangian in a regular prescribed way, 2)~tensor under arbitrary 
coordinate transformations, 3) symmetric in its components, 4) conserved due 
to the equations of motion derived from the same Lagrangian, 5) free of the 
second (highest) derivatives of the field variables, and 6) is unique up to 
trivial modifications not containing the field variables. There is nothing  
else, in addition to these 6 conditions, that one could demand from an 
energy-momentum object, acceptable both on physical and mathematical grounds.
The derived gravitational energy-momentum tensor should be useful in practical
applications. 
\end{abstract}
\vspace{1cm}

\newpage   
\section{Introduction}

      The notions of energy and momentum play important role in 
physics \cite{MTW},\cite{LL}.
These quantities are useful because they are conserved. The conservation
laws follow from the equations of motion, but we can gain important 
information about the system even without explicitely solving its equations 
of motion. 

	For a distributed system (or a field) the densities of energy, 
momentum, and flux of momentum are functions of points labelled by some 
coordinates $x^{\alpha}$. These functions combine in the  
energy-momentum tensor $T^{\mu\nu}(x^{\alpha})$, that is, the 
components of $T^{\mu\nu}$ transform according to the tensor rule under 
arbitrary transformations of the coordinates  $x^{\alpha}$
(independently of whether the space of points $x^{\alpha}$ is endowed  
with one or another metric tensor). It would be embarrassing to use an 
energy-momentum object which did not transform as a tensor under, say,
a transition from rectangular to spherical coordinates. Usually,
the $T^{\mu\nu}$ is a symmetric tensor, $T^{\mu\nu}=T^{\nu\mu}$.
The symmetry of $T^{\mu\nu}$ is required for a proper formulation of the
angular momentum conservation. The local distributions of
$T^{\mu\nu}(x^{\alpha})$ are important not only because they prescribe 
some numerical values to the energetic characteristics of the field, 
but also because they
can be viewed responsible for the local state of motion of particles 
and bodies interacting with the field.
In field theories governed by second-order differential
equations, one expects the energy-momentum tensor to depend on squares
of first-order derivatives of the field variables, but not on second 
derivatives.

	For Lagrangian-based theories, the derivation of the conserved 
energy-momentum object is closely related to the variational 
procedure by which the equations of motion are being derived (see, for
example, \cite{LL}). At the beginning it is better to speak about an 
energy-momentum object, rather than a tensor, because at the first steps
of derivation the 
transformation properties are either not being discussed or not obvious. 
In fact, there are two routes of derivation. One produces a
``canonical'' object, and another produces a ``metrical'' object.
The first route takes its origin from Euler and Lagrange.
This route does not care about
transformation  properties of the field variables and Lagrangian itself, and
whether the Lagrangian includes any metric tensor.
But what is important is whether the Lagrangian contains explicitely
(in a manner other than through the field variables) the independent
variables (coordinates) $x^{\alpha}$. If such dependence on $x^{\alpha}$ is 
present, one should not expect first integrals of the equations of
motion and conserved quantities. If there is no such a 
dependence, some sort of conservation laws is guaranteed as a consequence 
of the equations of motion.

	The second route is associated with the Noether identities. Here one 
exploits from the very beginning the transformation properties of
fields and Lagrangians. One requires the action to be a quantity independent
of any coordinate transformations and, hence, one requires the Lagrangian to
be a scalar density, that is, a scalar function times the square root of the
metric determinant. This route produces a ``metrical'' object, which is
essentially the variational derivative of the Lagrangian with respect to the
metric tensor. This object is automatically a symmetric 
tensor, and it is conserved if the equations of motion are satisfied. The
conserved tensors are usually understood in the sense that they 
obey differential conservation equations, but 
one can also derive from them the integral conserved 
quantities if, as is always required, the system is isolated.
For radiating systems, the fluxes of energy and momentum participate in the
balance equations.

       Both objects, canonical and metrical, are defined up to certain 
additive terms which do not violate equations of motion. These terms are 
a generalisation of the additive constant which arises even in a simplest
one-dimensional mechanical problem, when the Lagrangian does not depend 
on time explicitely. It is known that the first integral of 
the equation of motion,
which we interpret as energy, can be shifted by a constant. In field 
theories, the additive terms can be used to our advantage. For instance,
the canonical object can be made symmetric, if it was not such originally,
and the metrical object can be made free of second derivatives, if it
contained them originally. Despite the different routes of derivation, the
canonical and metrical objects are deeply related. If they are derived
from the same Lagrangian, explicitely containing metric tensor in
addition to field variables, they are equal to each other, up to a 
certain well defined expression calculable from the Lagrangian.

	In traditional field theories, one arrives, after some work, 
at the energy-momentum object which is: 1) derivable from the 
Lagrangian in a regular prescribed way, 2) a tensor
under arbitrary coordinate transformations, 3) symmetric in its 
components, 4) conserved due to the equations of motion obtained from the
same Lagrangian, 5) free of the second (highest) derivatives of the
field variables, and 6) is unique up to trivial modifications not containing 
the field variables.  There is nothing else, in addition to these 6
conditions, that we could demand from an acceptable energy-momentum
object, both on physical and mathematical grounds.

	When it comes to the gravitational field, as described by the
geometrical formulation of the general relativity, the things become more
complicated. It is often argued that the equivalence principle forbids 
gravitational energy-momentum tensor. What is meant in practice is that  
the all first derivatives of any metric tensor $g_{\mu\nu}(x^{\alpha})$ 
can be made,
by an appropriate choice of coordinates $x^{\alpha}$, equal to zero 
along the world line of a freely falling observer (along a 
timelike geodesic line). But the first derivatives of $g_{\mu\nu}(x^{\alpha})$
can be eliminated along any world line, not necessarily of a freely falling
observer. And this is true independently of the presence and form of coupling
of $g_{\mu\nu}(x^{\alpha})$ to other fields, and independently of whether the
$g_{\mu\nu}(x^{\alpha})$ obeys any equations. Since all components of a tensor
can not be eliminated by a coordinate transformation, this reference to a
physical principle is regarded to be an argument against a gravitational
energy-momentum tensor, but the argument sounds more like a fact from the
differential geometry. Despite of this argument, one usually notices that it
is desirable, nevertheless, to construct at least an ``effective" 
gravitational energy-momentum tensor. In practice, 
this means that we combine some terms of the Einstein equations, in one 
or another manner, into an object, which does not behave as a tensor even
under a transition from rectangular to spherical coordinates, but which 
possesses some desirable properties of the energy-momentum tensor, and this is
why it is an ``effective'' tensor. And, finally, one usually argues that
the ``effective'' tensor becomes the ``well-defined'' tensor after averaging
over several wavelengths. Obviously, this transmutation of a pseudotensor into
a tensor can be done only in an approximate and restricted sense. And, 
in general, the averaging over several wavelengths means that the 
numerical result will depend on whether we have averaged over, 
say, 3 or 30 wavelengths.

	This shaky situation can be tolerated as long as we are 
interested only in solving the Einstein equations. But this situation 
becomes risky when we need to know something more. It appears that the problem
of a rigorously defined energy-momentum tensor may have more than a purely
academic interest. We have in mind a specific question which was actually one
of motivations for our renewed interest to this problem.

	It is likely that the observed \cite{3} large-angular-scale
anisotropies in the microwave background radiation are caused by 
cosmological perturbations of quantum-mechanical origin. Cosmological
perturbations can be either purely gravitational fields, as in the case of
gravitational waves, or should necessarily involve gravitational component, 
as in the case of density perturbations.   
To make reliable theoretical predictions one needs to normalize 
the initial quantum fluctuations. In words, this means to assign energy of
a half of the quantum to each mode. In practice, this implies 
the availability of a rigorously defined energy-momentum tensor for the 
field in question, which allows to enforce the energy 
$\frac1{2} \hbar \omega$,
and not, say, $\frac 1{3} \hbar \omega$ or $30 \hbar \omega$, 
for the initial quantum
state. A change in the numerical coefficient would lead to
the corresponding change in the final results. The preliminary calculations
show that the contributions of the quantum-mechanically produced 
gravitational waves and density perturbations should be approximately
equal, with some preference to gravitational waves \cite{LPG}. A detailed 
analysis of the available observational data \cite{MSV} seems to favour the 
gravitational wave contribution twice as large as that of density 
perturbations. Remarkably, the factor of 2 may turn out to be important when 
comparing the theoretical predictions with observations. This is why, in
our opinion, we cannot afford even a numerical coefficient ambiguity in such 
fundamental constructions as gravitational energy-momentum tensor.

	We believe that the difficulty in deriving a proper 
gravitational energy-momentum tensor lies in the way we treat gravity,
not in the nature of gravity as such. In the geometrical formulation
of the general relativity, the components $g_{\mu\nu}(x^{\alpha})$ play a
dual role. From one side they are components of the metric tensor, from the
other side they are considered gravitational field variables. If one insists
on the proposition that ``gravity is geometry'' and ``geometry is gravity'',
then, indeed, it is impossible to derive from the Hilbert-Einstein Lagrangian
something reasonable, satisfying the 6 conditions listed above. But the
geometrical approach to the general relativity is not the only one available.
It is here where it is necessary to look at the general relativity from 
the field-theoretical positions. The general relativity can be perfectly well
formulated as a strict non-linear field theory in flat space-time. This is 
a different formulation of the theory, not a different theory. The
importance of looking at theories from different viewpoints
was well emphasized by Feynman \cite{Feynman}: ``if the peculiar viewpoint 
taken
is truly experimentally equivalent to the usual in the realm of the 
known there is always a range of applications and problems in this realm
for which the special viewpoint gives one a special power and clarity 
of thought, which is valuable in itself''.

	The field-theoretical formulation of the general relativity
treats gravity as a non-linear tensor field $h^{\mu\nu}(x^{\alpha})$ in the
Minkowski space-time. In arbitrary curvilinear coordinates, the metric tensor
of the flat space-time is $\gamma_{\mu\nu}(x^{\alpha})$. If necessary, one is
free to use the Lorentzian coordinates and to transform
$\gamma_{\mu\nu}(x^{\alpha})$ into the usual constant matrix $\eta_{\mu\nu}$.
The Minkowski metric is not an artificially imposed ``prior geometry'', but a
reflection of experimental facts. We know that far away from gravitating
bodies, and whenever gravitational field can be neglected, the 
space and time intervals satisfy the requirements of the
Minkowski space-time. In the presence of the gravitational field, all
kinds of ``rods'' and ``clocks'' will exhibit violations of the Minkowski 
relationships. This is a result of the universality of the gravitational
interaction (as we understand it today). One is free to interpret the 
results of the measurement as a manifestation of the curvature of the 
space-time, rather than the action of the universal gravitational 
field. In this sense, the Minkowski space-time
becomes ``unobservable''. But this does not mean that the Minkowski metric
is illegitimate or useless. On the contrary, it is being routinely
used in relativistic astrometry and relativistic celestial mechanics.
People are well aware of the general relativity and curved space-time. 
But it turns out to be more convenient and informative to store and 
analyze the data in terms of the ``unobservable'' flat space-time 
quantities (after subtraction of the theoretically
calculated general-relativistic corrections), rather than in terms of
directly measured ``observable" quantities. If this is possible and useful
in the regime of weak gravitational fields, it can be useful for any
fields. In fact, for the problem of the
gravitational energy-momentum tensor, the use of the Minkowski metric
allows one to put everything in full order. The demonstration of 
this fact is the main purpose of the paper.            
	
The structure of the paper is as follows. 

In Sec. II we review definitions of the canonical and metrical 
energy-momentum tensors for general field theories. The associated 
ambiguities, and their relationship with the equations of
motion, is a considerable technical complication on its own. However, we
show in detail how the canonical and metrical tensors are related. The main
conclusion is that, whatever the starting point, the allowed adjustments lead 
eventually to one and the same object satisfying the imposed requirements.   
We use this general analysis in Sec. IV in course of derivation of the  
gravitational energy-momentum tensor. Sec. III is devoted to the 
field-theoretical formulation of the general relativity.
We start from the case of pure gravity, without matter sources. 
The gravitational Lagrangian and field equations are given explicitely.  
It is shown that the derived field equations, plus their appropriate
interpretation, are fully equivalent to
the Einstein equations in the geometrical formulation. In Sec. IV, being
armed with the gravitational Lagrangian and field equations, we apply
the general definitions of Sec. II for derivation of the gravitational
energy-momentum tensor. By different routes we arrive at the 
energy-momentum tensor satisfying all 6 demands listed in the Abstract
of the paper. It is shown that this tensor is unique up to trivial 
modifications which do not involve the field variables. We call this 
object the true energy-momentum tensor.  
In Sec. V we analyze the way in which the true energy-momentum tensor
participates in the non-linear gravitational field equations. 
The gravitational
energy-momentum tensor is not, and should not be, a source in the 
``right-hand side of Einstein's equations". But it is a source for
the generalised (non-linear) d'Alembert operator. It is shown that a
geometrical object most closely related to the derived energy-momentum
tensor is the Landau-Lifshitz pseudotensor. Their numerical values (but
not the transformation properties) are equal at least under some
conditions. In Sec. VI we include matter fields in our consideration
and define energy-momentum tensor for the matter fields. The
gravitational energy-momentum tensor is now modified because of the 
presence of the matter Lagrangian. However, both, gravitational and
matter energy-momentum tensors participate in the gravitational
field equations at the equal footing. Their sum is the total energy-momentum
tensor which is now the source for the previously mentioned 
generalised (non-linear) d'Alembert operator.    
The conservation laws for the total energy-momentum tensor are
guaranteed by general theorems (Sec. II) and are manifestly 
satisfied as a differential consequence of the field equations. The
derived equations, plus their appropriate interpretation, are fully 
equivalent to the Einstein's geometrical equations with matter. The
final Sec.~VII contains conclusions. Some technical details are relegated 
to Appendix A and Appendix B.

\section{Definitions of the Energy-Momentum Tensor}

  Some of the  material of this section is known in the literature but we
  present it in a systematic way and in a form appropriate for our
  further treatment of the general relativity as a field theory in
  flat space-time.

\subsection{The canonical energy-momentum tensor}

Let us first recall how the notion of energy arises in the simplest case
of a 1-dimensional mechanical system
with the Lagrangian $L=L(q,\dot q, t)$ and the  action
$$
S= \int_{t_1}^{t_2} L(q, \dot q,t) \; dt~. \nonumber
$$
The equation of motion (the Euler - Lagrange equation) follows from the
requirement that the action is stationary, $\delta S = 0$, under arbitrary 
variations of $q(t)$ vanishing at the limits of integration (what we will
always assume):
\bea
\frac{\partial L}{\partial q} - \frac{d}{dt}
\left( \frac{\partial L}{\partial \dot q}\right) =0~. \label{1dim}
\ena
The symbol of the total derivative $d/dt$ emphasizes the need to
include the partial derivative
by $t$ if the function $L$ depends on time explicitely.  
If the Lagrangian does not depend on time $t$ explicitely, the equation
(\ref{1dim}) admits the first integral. In this case one has
\bea
\frac {dL}{dt} = \frac{\partial L}{\partial q} \dot q  +
\frac{\partial L}{\partial \dot q} \ddot q  
\;\;\:{\rm and}\;\;\:
\frac{d}{dt}\left( \frac{\partial L}{\partial \dot q} \right)=
\frac{\partial}{\partial q} \left( \frac{\partial L}{\partial \dot q} \right)
\dot q  + 
\frac{\partial}{\partial \dot q} \left( \frac{\partial L}{\partial \dot q}\right)
\ddot q~. \label{02}
\ena
By multiplying eq. (\ref{1dim}) with $\dot q$ and rearranging the terms
with the help of (\ref{02}), one transforms eq. (\ref{1dim}) to
\bea
\frac{d}{dt} \left(\dot q  \frac{\partial L}{\partial \dot q} - L\right) =0~.
\nonumber
\ena
This equation has the form of a conservation law, and the quantity
$
E =\dot q \frac{\partial L}{\partial \dot q} - L, \nonumber
$
called energy, is a constant. With the same success we could call energy
the quantity  
$
E = \left(\dot q \frac{\partial L}{\partial \dot q} - L \right) + C, \nonumber
$
where $C$ is a constant. The equation of motion (\ref{1dim}) is still satisfied. 

     These considerations apply to any field theory described by 
the Lagrangian $L=L(q_A;q_{A,\alpha}; x^{\alpha})$ where $q_A(x^{\alpha})$
is a set of variables, and $x^{\alpha}$ is a set of coordinates.
The variational principle produces the field equations
\bea
\frac{\partial L}{\partial q_A}  -
\left( \frac{\partial L}{\partial q_{A ,\alpha}}\right)_{,\alpha} = 0, 
\label{eqm}
\ena
where the last differentiation with respect to $x^{\alpha}$ includes 
the partial derivative by $x^{\alpha}$, and the summation over repeated
indices is (always) assumed. The field equations are conveniently written
as
$
\frac{\delta L}{\delta q_A}=0,\nonumber
$
where the variational derivative $\delta/ \delta$ denotes (see, for example,
\cite{DeW}) 
\bea
\frac{\delta L(q_A; q_{A ,\alpha}; x^{\alpha})}
{\delta q_A} \equiv
 \frac{\partial L}{\partial q_A}  -
\left( \frac{\partial L}{\partial q_{A ,\alpha}}\right)_{,\alpha}~.
\label{varder}
\ena
If Lagrangian depends on second derivatives, the right-hand-side of
(\ref{varder}) acquires an extra term, see Appendix B.

	If the function $L$ does not depend 
on $x^{\alpha}$ explicitely, one expects that the field equations can
be transformed into the conservation equations, 
equal in number to the number of coordinates $x^{\alpha}$. In this case,
one has  
\bea
L_{,\sigma} = 
\frac{\partial L}{\partial q_A} q_{A ,\sigma} +
\frac{\partial L}{\partial q_{A ,\tau}} q_{A ,\tau ,\sigma}~. \nonumber
\ena
By multiplying eq. (\ref{eqm}) with $q_{A,\sigma}$, taking summation over
$A$, and making rearrangements similar to the ones described above, one
obtains, as a consequence of the field equations, 
\bea
\left(q_{A, \alpha} \frac{\partial L}{\partial q_{A ,\beta}}
- \delta^{\beta}_{\alpha} L \right)_{, \beta} =0~. \nonumber
\ena
The expression
\bea
\stackrel{c}{t}\!_{\alpha}^{\ \beta} = 
q_{A, \alpha} \frac{\partial L}{\partial q_{A ,\beta}} -
\delta^{\beta}_{\alpha} L \nonumber
\ena
is the canonical (label $c$) conserved energy-momentum object. The upper
or lower positions of $\alpha$ and $\beta$ are not essential, but the 
positions of the first index ($\alpha$) and the second index ($\beta$) 
are distinguishable. In general, the object  
$\stackrel{c}{t}\!_{\alpha}^{\ \beta}$ is not symmetric 
in $\alpha$ and $\beta$. 

       With the same success we could write for the canonical object
\bea
\stackrel{c}{t}\!_{\alpha}^{\ \beta} = q_{A,\alpha}
\frac {\partial L}{\partial q_{A,\beta}}
-\delta^{\beta}_{\alpha}L + \Psi_{\alpha}^{\ \beta}, \nonumber
\ena
if the function $\Psi_{\alpha}^{\ \beta}$ satisfies
\bea
{\Psi_{\alpha}^{\ \beta}}_{,\beta}=0 \label{ct55}
\ena
identically or due to the equations of motion (\ref{eqm}). 
In order to satisfy  
(\ref{ct55}) identically, it is sufficient to have 
$\Psi_{\alpha}^{\ \beta}={\psi_{\alpha}^{\ \beta\tau}}_{,\tau}$ where
$\psi_{\alpha}^{\ \beta\tau}$ is antisymmetric in $\beta$ and $\tau$: 
$
\psi_{\alpha}^{\ \beta\tau} = - \psi_{\alpha}^{\ \tau\beta},\nonumber
$
so that ${\psi_{\alpha}^{\ \beta\tau}}_{,\tau,\beta} \equiv 0$. The 
function $\psi_{\alpha}^{\ \beta\tau}$ is usually called a superpotential.
By an appropriate choice of 
$\Psi_{\alpha}^{\ \beta}$ one can make the object  
$\stackrel{c}{t}\!_{\alpha}^{\ \beta}$ symmetric in its components. 
The transformation properties of 
$\stackrel{c}{t}\!_{\alpha}^{\ \beta}$ under coordinate transformations
are not defined until the transformation properties of the field variables
and $L$ are defined.   

	We now move to covariant relativistic theories. One normally considers
physical fields of various tensor ranks (scalar, vector, tensor, etc.)
in a space-time with some metric tensor. The Lagrangian $L$ is required to be
a scalar density with respect to arbitrary coordinate transformations, 
that is, $L$ is a scalar function times the square root of the (minus) 
metric determinant. For a better contact with our further study, we  
consider a symmetric tensor field $h^{\mu\nu}(x^{\alpha})$ placed in a flat
space-time with the metric tensor $\gamma_{\mu \nu}(x^{\alpha})$ written
in arbitrary curvilinear coordinates $x^{\alpha}$. The general form for 
the Lagrangian density is    
\bea
L=L(\gamma^{\mu\nu}, h^{\mu\nu}, {h^{\mu\nu}}_{;\beta}) \label{L*}
\ena
where ``;'' denotes a covariant derivative defined by $\gamma_{\mu \nu}$
and the associated connection (Christoffel symbols) 
$C^{\alpha}_{\ \mu\nu}$. 
The $\gamma^{\mu\nu}$ and $C^{\alpha}_{\ \mu\nu}$ 
are functions of $x^{\alpha}$ but they are not dynamical
variables, and hence they make the $L$
dependent on $x^{\alpha}$ explicitely. On the general grounds, one does not 
expect the Euler-Lagrange equations to reduce to any conservation  
equations in the usual sense, i.e. in terms of vanishing partial 
derivatives. However, since 
${\gamma^{\mu\nu}}_{;\alpha} \equiv 0$, one can derive a covariant 
generalisation of the conservation laws, i.e. in terms of vanishing
covariant derivatives. This is, of course, consistent with our ability
to choose coordinates $x^{\alpha}$ in such a way that
$\gamma^{\mu\nu}$ will become a constant matrix and 
$C^{\alpha}_{\ \mu\nu}$ will all vanish, thus removing the explicit
dependence of $L$ on coordinates. Moreover, as we will show below, 
the vanishing covariant divergence will apply to the canonical 
energy-momentum tensor, which is now a manifestly tensorial quantity.
 
	Let us first give a covariant generalisation to the equations
of motion. The action for the Lagrangian (\ref{L*}) is
\bea
S= \frac 1{c} \int L \; d^4 x~,  \label{a0}
\ena
where the integral is taken over some 4-volume $V$.  
Considering $\delta h^{\mu\nu}$ and $\delta {h^{\mu\nu}}_{ ;\alpha}$ 
as independent variations we can write:  
\bea
\delta L = \frac{\partial L}{\partial h^{\mu\nu}} \delta h^{\mu\nu}  +
 \frac{\partial L}{\partial {h^{\mu\nu}}_{ ;\tau}}\delta {h^{\mu\nu}}_{ ;\tau}~. 
\label{a1}
\ena
It is easy to check that the operations of variation and covariant 
differentiation commute.   
Using this property in (\ref{a1}) we have 
\bea
\delta L = \frac{\partial L}{\partial h^{\mu\nu}} \delta h^{\mu\nu}  +
\left(\frac{\partial L}{\partial {h^{\mu\nu}}_{ ;\tau}}
\delta {h^{\mu\nu}}_{}\right)_{ ;\tau} -
\left(\frac{\partial L}{\partial {h^{\mu\nu}}_{ ;\tau}}\right)_{;\tau}
\delta {h^{\mu\nu}}_{}~.
\label{a3}
\ena
Since the quantity 
$\frac{\partial L}{\partial {h^{\mu\nu}}_{ ;\tau}} \delta h^{\mu\nu}$ 
is a vector density of weight 1 (i.e. a vector quantity 
times~$\sqrt{-\gamma}$) we have
\bea
\left(\frac{\partial L}{\partial {h^{\mu\nu}}_{ ;\tau}}\delta h^{\mu\nu}\right)_{ 
;\tau} =
\left(\frac{\partial L}{\partial {h^{\mu\nu}}_{ ;\tau}}\delta h^{\mu\nu}\right)_{ 
,\tau}~.\label{a33}
\ena
Substituting (\ref{a3}) into (\ref{a0}) and taking into account the
equality above one obtains
\bea
\delta S = \frac 1{c} \int
\left[  \frac{\partial L}{\partial h^{\mu\nu}} \delta h^{\mu\nu} 
 - \left(\frac{\partial L}{\partial {h^{\mu\nu}}_{ ;\tau}}\right)_{;\tau}\delta 
{h^{\mu\nu}}_{}+
\left(\frac{\partial L}{\partial {h^{\mu\nu}}_{ ;\tau}}\delta 
{h^{\mu\nu}}_{}\right)_{ ,\tau}
\right]\; d^4x =0~. \label{a4}
\ena
At the boundary of integration we have $\delta h^{\mu\nu}=0$, so the
integral of the last term in (\ref{a4}) is zero.  
The variations $\delta {h^{\mu\nu}}_{}$ are arbitrary, and we arrive at
the field equations in an explicitely covariant form: 
\bea
\frac{\partial L}{\partial h^{\mu\nu}}-
\left(\frac{\partial L}{\partial {h^{\mu\nu}}_{ ;\tau}}\right)_{;\tau} =0~.
\label{cfe}
\ena
Certainly, one could have obtained the same result in a more familiar way,
starting from the Lagrangian in the form containing $h^{\mu\nu}$ and
the ordinary (rather than covariant) derivatives
${h^{\mu\nu}}_{,\tau}$ (see Appendix A).

	One can now derive the canonical energy-momentum object in  
exactly the same way as was described before. 
Namely, one multiplies the field equations (\ref{cfe}) by 
${h^{\mu\nu}}_{;\sigma}$ and rearranges the terms to arrive
at the covariant conservation law:
\bea
\left({h^{\mu\nu}}_{;\alpha}\frac {\partial L}{\partial {h^{\mu\nu}}_{;\beta}}
- \delta^{\beta}_{\alpha}L\right)_{;\beta} =0~.\nonumber
\ena
The expression 
\bea
\stackrel{c}{t}\!^{\alpha \beta} =
\frac 1{\sqrt{-\gamma}}\left(\gamma^{\alpha \tau} {h^{\mu\nu}}_{;\tau}
\frac {\partial L}{\partial {h^{\mu\nu}}_{;\beta}} -
\gamma^{\alpha \beta}L\right)              \label{cct}
\ena
is the canonical energy-momentum tensor. We could also define  
$\stackrel{c}{t}\!^{\alpha\beta}$ as 
\bea
\stackrel{c}{t}\!^{\alpha \beta} =
\frac 1{\sqrt{-\gamma}}\left(\gamma^{\alpha \tau} {h^{\mu\nu}}_{;\tau}
\frac {\partial L}{\partial {h^{\mu\nu}}_{;\beta}} -
\gamma^{\alpha \beta}L\right) + \Psi^{\alpha\beta}, \nonumber
\ena
where $\Psi^{\alpha\beta}$ is a function  
such that ${\Psi^{\alpha\beta}}_{;\beta}=0$, identically or  
due to the field equations. The conserved canonical energy-momentum object 
(\ref{cct}) does not contain second order derivatives and is 
manifestly a tensorial quantity, but, in general, the canonical
energy-momentum tensor is 
not symmetric in its components. However, it can be made symmetric by 
an appropriate choice of a non-symmetric $\Psi^{\alpha\beta}$ .

\subsection{The metrical energy-momentum tensor}

	From the general Lagrangian (\ref{L*}) one can also derive the metrical
energy-momentum tensor. Its derivation relies on the transformation
properties of all the participating quantities with respect to 
coordinate transformations. 
	
	An infinitesimal coordinate transformation
\bea
\tilde x^{\alpha} = x^{\alpha} - \xi^{\alpha}(x^{\beta}) \label{ctr}
\ena
generates the Lie transformations along the vector field $\xi^{\alpha}$, 
which can be presented as corresponding
variations of the field variables, of the metric tensor, and of the 
Lagrangian:  
$\delta h^{\mu\nu}$, $\delta \gamma^{\mu\nu}$, and $\delta L$, respectively.
Since the Lagrangian (\ref{L*}) is a scalar density, its variation is a
total derivative 
\bea
\delta L =  (L \xi^{\alpha})_{,\alpha}~. \label{mt1}
\ena
The change in the metric tensor is
\bea
\delta\gamma^{\mu\nu}=- \xi^{\mu;\nu} - \xi^{\nu;\mu}~. \label{lim}
\ena 
And there is also a corresponding change in the field variables 
\bea
\delta h^{\alpha\beta} = 
\xi^{\sigma}{h^{\alpha\beta}}_{;\sigma}
- h^{\alpha\sigma}{\xi^{\beta}}_{;\sigma} -
h^{\beta\sigma}{\xi^{\alpha}}_{;\sigma} \label{lifi}
\ena
but we will not need to know its concrete form for this derivation.    
	
	Taking into account (\ref{mt1}) and assuming that the vector field
$\xi^{\alpha}$ vanishes at the boundary of integration, we conclude	 
that the variation of the action must be equal to zero:  
\bea
\delta S = \frac{1}{c} \int \delta L \; d^4 x = 0~. \label{dS0}
\ena
On the other hand, we know that an arbitrary variation of $L$, not
necessarily caused by (\ref{ctr}), has the general form 
\bea
\delta L = \frac{\delta L}{\delta h^{\mu\nu}} \delta h^{\mu\nu} + 
 \frac{\delta L}{\delta \gamma^{\mu\nu}} \delta \gamma^{\mu\nu}+ 
 {A^{\alpha}}_{,\alpha}  \label{M1}
\ena
where
\bea
A^{\alpha} =
\frac{\partial L}{\partial {h^{\mu\nu}}_{,\alpha}} \delta h^{\mu\nu} + 
\frac{\partial L}{\partial {\gamma^{\mu\nu}}_{,\alpha}} 
\delta \gamma^{\mu\nu}~. \nonumber  
\ena
In writing this formula we took into account the fact that the
first derivatives of $\gamma^{\mu \nu}$  
participate, through the Christoffel symbols, in the covariant derivatives 
of the field variables.

	At this point we require that the field equations are satisfied
\bea
\frac{\delta L}{\delta h^{\mu\nu}} = 0 \label{Mem}
\ena
so the first term in (\ref{M1}) is zero. In the second term of (\ref{M1})
we use the specific variation (\ref{lim}). Then, eq. (\ref{dS0}) acquires
the form 
\bea
-2\int \left(
\frac{\delta L}
{\delta \gamma_{\rho\sigma}} \right)_{;\sigma} 
\xi_{\rho}\; d^4 x
+ \int \left( 2 \frac{\delta L}
{\delta \gamma_{\alpha\beta}}
 \xi_{\beta}+ A^{\alpha}\right)_{,\alpha} \; d^4 x =0 \label{new} 
\ena
where we have also used the following equality:
\be
\frac{\partial L}{\partial \gamma^{\alpha\beta}}=
- \gamma_{\mu\alpha}\gamma_{\nu\beta}
\frac{\partial L}{\partial \gamma_{\mu\nu}}~. \label{tuptd}
\en
The second integral in (\ref{new}) transforms into a surface 
integral and vanishes under appropriate boundary conditions for 
$\xi^{\alpha}$. 
Since the functions $\xi_{\rho}(x^{\alpha})$ are arbitrary, we finally
obtain:
\bea
-2\left( \frac{\delta L}{\delta \gamma_{\rho\sigma}}\right)_{;\sigma} =0~.
\label{dLdg}
\ena

	The metrical (symbol $m$) energy-momentum tensor
$\stackrel{m}{t}\!^{\mu\nu}$ is defined as 
\bea
\stackrel{m}{t}\!^{\mu\nu} =
-\frac{2}{\sqrt{-\gamma}}\frac{\delta L}{\delta \gamma_{\mu\nu}}~,  \label{mt}
\ena
so that eq. (\ref{dLdg}) takes the form of the covariant conservation law
(valid only on solutions to the equations of motion (\ref{Mem})): 
\bea
{\stackrel{m}{t}\!^{\mu\nu}}_{;\nu}= 0~. \label{mcon} 
\ena
As before, one can also write for the metrical energy-momentum tensor
\bea
\stackrel{m}{t}\!^{\mu\nu} =
-\frac{2}{\sqrt{-\gamma}}\frac{\delta L}{\delta \gamma_{\mu\nu}} +
\Phi^{\mu\nu}~, \label{mt'} 
\ena
where the function $\Phi^{\mu\nu}$ satisfies
$
{\Phi^{\mu\nu}}_{;\nu}= 0
$
identically or due to the field equations. The derived conserved object
(\ref{mt}) is automatically symmetric and a tensor, but, as a rule, it 
contains second order 
derivatives of the field variables, even if the Lagrangian does not 
contain them. They are generated by one extra differentiation in the 
definition of the variational derivative (see the second term in eq.
(\ref{varder})).   However, by an appropriate use of 
$\Phi^{\mu\nu}$ and the field equations, all second derivatives can be
removed, as we will discuss in detail later on.  
	
It is important to note that nothing in the derivation of eqs.
(\ref{mt}), (\ref{mcon}) actually required the $\gamma^{\mu \nu}$ to be a
metric tensor of a flat space-time,
that is, to have the curvature tensor constructed from $\gamma^{\mu \nu}$
equal to zero. One can still formally arrive at the equation, similar in
structure to eq. (\ref{mcon}), in arbitrary curved ``background''
space-time, where the covariant derivatives are now being taken with respect
to the curved metric. 
This is an element of a field theory in the ``background'' space-time, which 
is useful in some applications (see, for example, \cite{GPP},
\cite{PK}). As soon 
as
the field equations are satisfied, the corresponding covariant ``conservation
laws'' must be valid. However,
in this case, there is not and there should not be, in general, any 
conservation laws in the usual sense. First, one normally encounters severe
integrability conditions for the field equations. The number of independent
solutions, in the sense of the Cauchy problem, can be diminished, or the
solutions may not exist at all. Second,  
the vanishing covariant divergence cannot be converted into the
vanishing ordinary divergence. This is a well known formal obstacle,
but it has deep
and clear physical reasons: the ``background'' space-time is by itself
a gravitational field which interacts with a system and can exchange energy
with the system. For instance, even in the simplest Friedman-Robertson-Walker
space-times, gravitational waves can be amplified and gravitons can be
created \cite{10}.

	Returning to the strictly defined energy-momentum tensors, we
will now show that the canonical tensor (\ref{cct}) and the
metrical tensor (\ref{mt}) are closely related.

\subsection{Connection between metrical and canonical tensors}

	The metrical tensor (\ref{mt}) and the canonical 
tensor (\ref{cct}) are derived from the same Lagrangian (\ref{L*}), so one expects
them to be related. To find the link between $\stackrel{m}{t}\!^{\mu\nu}$ and 
$\stackrel{c}{t}\!^{\mu \nu}$ we return to the derivation of  
$\stackrel{m}{t}\!^{\mu\nu}$ based on the infinitesimal transformation
(\ref{ctr}).  

	It is convenient to write the variation (\ref{mt1}) in the form
\bea
\delta L = (L \xi^{\alpha})_{;\alpha}~. \label{dL}
\ena
The replacement of the ordinary divergence by the covariant one is 
allowed, because the differentiated quantity $(L \xi^{\alpha})$ 
is a vector density. We can also write
the general variation (\ref{M1}) in the form  
\bea
\delta L = \frac{\delta L}{\delta h^{\alpha\beta}} \delta h^{\alpha\beta} + 
 \frac{\delta L}{\delta \gamma^{\mu\nu}} \delta \gamma^{\mu\nu}+
 \left(\frac{\partial L}{\partial {h^{\alpha\beta}}_{;\tau}}
 \delta h^{\alpha\beta} \right)_{;\tau}+
 \left( \frac{\partial L}{\partial {\gamma^{\mu\nu}}_{,\tau}}
 \delta \gamma^{\mu\nu}\right)_{,\tau}~. \label{N1}
\ena
In writing this expression we took into account (\ref{a33}) and the fact that
\be
\frac{\partial L}{\partial {h^{\alpha\beta}}_{;\tau}}  = 
\frac{\partial L}{\partial {h^{\alpha\beta}}_{,\tau}}~.  \nonumber
\en
We will now show that the differentiated quantity in the last term of (\ref{N1}) 
is also a vector density, so that the ordinary divergence can be replaced
by the covariant one. Indeed, from the structure of (\ref{L*}) it follows that
\bea
\frac{\partial L}{\partial {\gamma^{\mu\nu}}_{,\tau}} =
 \frac{\partial L}{\partial {h^{\alpha\beta}}_{;\rho}}
 \frac{\partial {h^{\alpha\beta}}_{;\rho}}{\partial{\gamma^{\mu\nu}}_{,\tau}}  =
2\frac{\partial L}{\partial {h^{\alpha\beta}}_{;\rho}}h^{\sigma\beta}
 \frac{\partial C^{\alpha}_{\ \sigma\rho}}
 {\partial {\gamma^{\mu\nu}}_{,\tau}}~.    \nonumber
\ena
Since
\bea
\frac{\partial C^{\tau}_{\ \lambda\rho}}{\partial 
{\gamma^{\alpha\beta}}_{,\omega}} &=&
-\frac1{4}(\delta^{\omega}_{\rho}\delta^{\tau}_{\alpha}\gamma_{\beta\lambda}
+ \delta^{\omega}_{\rho}\delta^{\tau}_{\beta}\gamma_{\alpha\lambda}+
\delta^{\omega}_{\lambda}\delta^{\tau}_{\alpha}\gamma_{\beta\rho}+
\delta^{\omega}_{\lambda}\delta^{\tau}_{\beta}\gamma_{\alpha\rho}-
\gamma^{\tau\omega}\gamma_{\rho\alpha}\gamma_{\beta\lambda}-
\gamma^{\tau\omega}\gamma_{\rho\beta}\gamma_{\alpha\lambda}) \label{2c9}
\ena
it is now clear that the quantity  $
\frac{\partial L}{\partial {\gamma^{\mu\nu}}_{,\tau}}
 \delta \gamma^{\mu\nu}
$ is a vector density. Thus, we can rewrite (\ref{N1}) as   
\bea
\delta L = \frac{\delta L}{\delta h^{\alpha\beta}} \delta h^{\alpha\beta} + 
 \frac{\delta L}{\delta \gamma^{\mu\nu}} \delta \gamma^{\mu\nu}+
 \left(2\frac{\partial L}{\partial {h^{\alpha\beta}}_{;\rho}}h^{\sigma\beta}
 \frac{\partial C^{\alpha}_{\ \sigma\rho}}
 {\partial {\gamma^{\mu\nu}}_{,\tau}} \delta \gamma^{\mu\nu} +
 \frac{\partial L}{\partial {h^{\alpha\beta}}_{;\tau}}
 \delta h^{\alpha\beta} \right)_{;\tau}~. \label{N2}
\ena

	The expression (\ref{N2}) is valid for arbitrary variations, 
and hence it is valid for specific variations (\ref{lim}), (\ref{lifi})
caused by (\ref{ctr}). 
Therefore, the difference between (\ref{dL}) and (\ref{N2}) must be equal
to zero. Substituting (\ref{lifi}) and (\ref{lim}) into this difference, and
combining in separate groups the terms which contain
$\xi^{\sigma}$, ${\xi^{\sigma}}_{;\tau}$ and 
${\xi^{\sigma}}_{;\tau ;\lambda}$ one obtains the equality which
should be true for arbitrary vector field  
$\xi^{\sigma}(x^{\alpha})$ : 
\bea
\left[ \frac{\delta L}{\delta h^{\alpha\beta}}
 {h^{\alpha\beta}}_{;\sigma} +
 \left( \frac{\partial L}{\partial {h^{\alpha\beta}}_{;\tau}}
 {h^{\alpha\beta}}_{;\sigma} - \delta^{\tau}_{\sigma}L \right)_{;\tau}
\right]\xi^{\sigma} +  \nonumber \\
\left[ - 2\frac{\delta L}{\delta \gamma^{\rho\sigma}}\gamma^{\rho\tau} +
\left( \frac{\partial L}{\partial {h^{\alpha\beta}}_{;\tau}}
 {h^{\alpha\beta}}_{;\sigma} - \delta^{\tau}_{\sigma}L \right) -
2\frac{\delta L}{\delta h^{\alpha\sigma}}
 h^{\tau\alpha} + 
 \right.\nonumber \\
 \left.
 \left(-4 \frac{\partial L}{\partial {h^{\alpha\beta}}_{;\rho}}
 h^{\phi\beta} \frac{\partial C^{\alpha}_{\ \phi\rho}}
 {\partial {\gamma^{\sigma\nu}}_{,\lambda}} \gamma^{\nu\tau} -
2 \frac{\partial L}{\partial {h^{\alpha\sigma}}_{;\lambda}}
h^{\tau\alpha} \right)_{;\lambda}\right]{\xi^{\sigma}}_{;\tau} +
\nonumber\\
\left[-4 \frac{\partial L}{\partial {h^{\alpha\beta}}_{;\rho}}
 h^{\phi\beta} \frac{\partial C^{\alpha}_{\ \phi\rho}}
 {\partial {\gamma^{\sigma\nu}}_{,\lambda}} \gamma^{\nu\tau} -
2 \frac{\partial L}{\partial {h^{\alpha\sigma}}_{;\lambda}}
h^{\tau\alpha}
\right]{\xi^{\sigma}}_{;\lambda ;\tau} =0~. \label{N3} 
\ena

	The coefficient in front of $\xi^{\sigma}$ is identically zero,
because all the terms cancel out. 
To check this one has to recall the definition of the variational
derivative
$$
\frac{\delta L}{\delta h^{\alpha\beta}}=
\frac{\partial L}{\partial h^{\alpha\beta}} -
\left( \frac{\partial L}{\partial {h^{\alpha\beta}}_{,\tau}}
\right)_{,\tau} ~, 
$$
to use (\ref{d/d}) and (\ref{d/d,}) for
$\frac{\partial L}{\partial h^{\alpha\beta}}$ and
$\left(\frac{\partial L}{\partial {h^{\alpha\beta}}_{,\tau}}\right)_{,\tau}$,
 and to take into account
$$
\frac{\partial L}{\partial h^{\alpha\beta}} {h^{\alpha\beta}}_{ ;\sigma} +
\frac{\partial L}{\partial {h^{\alpha\beta}}_{ ;\tau}}
{h^{\alpha\beta}}_{ ;\tau ;\sigma} = L_{;\sigma}~.
$$
The last term in (\ref{N3}), which contains ${\xi^{\sigma}}_{;\lambda ;\tau}$,  
is also identically zero. This is true because the 
${\xi^{\sigma}}_{;\lambda ;\tau}$ is symmetric in the indices
$\lambda ,\tau$ whereas the coefficient is antisymmetric in these
indices. To show this in detail, we denote this coefficient  
$\sqrt{-\gamma} \psi_{\sigma}^{\ \tau\lambda}$ and rewrite it
using formula (\ref{2c9}): 
\bea
\sqrt{-\gamma} \psi_{\sigma}^{\ \tau\lambda} = 
-4 \frac{\partial L}{\partial {h^{\alpha\beta}}_{;\rho}}
 h^{\phi\beta} \frac{\partial C^{\alpha}_{\ \phi\rho}}
 {\partial {\gamma^{\sigma\nu}}_{,\lambda}} \gamma^{\nu\tau} -
2 \frac{\partial L}{\partial {h^{\alpha\sigma}}_{;\lambda}}
h^{\tau\alpha} =
 \nonumber \\
\left(\frac{\partial L}{\partial {h^{\sigma\beta}}_{;\tau}}h^{\lambda\beta} -
\frac{\partial L}{\partial {h^{\sigma\beta}}_{;\lambda}}h^{\tau\beta}\right)
 +
  \frac{\partial L}{\partial {h^{\alpha\beta}}_{;\phi}}
  (h^{\lambda\beta}\gamma^{\alpha\tau}-h^{\tau\beta} \gamma^{\alpha\lambda})
\gamma_{\phi\sigma}
 +
 \nonumber \\
\gamma_{\phi\sigma} h^{\phi\beta}
\left(\frac{\partial L}{\partial {h^{\alpha\beta}}_
{;\lambda}}  \gamma^{\alpha\tau} -
\frac{\partial L}{\partial {h^{\alpha\beta}}_{;\tau}}
 \gamma^{\alpha\lambda} \right)~. \label{psi0}
\ena
It is now clear that $\psi_{\sigma}^{\ \tau\lambda}=
-\psi_{\sigma}^{\ \lambda\tau}~$.
	So, we are left only with the term which contains  
${\xi^{\sigma}}_{ ;\tau}$. 
Since the vector field $\xi^{\sigma}$ is arbitrary, this gives us the
equation
\bea
- 2\frac{\delta L}{\delta \gamma^{\rho\sigma}}\gamma^{\rho\tau} +
\left( {h^{\alpha\beta}}_{;\sigma} 
\frac{\partial L}{\partial {h^{\alpha\beta}}_{;\tau}} - 
\delta^{\tau}_{\sigma}L \right) -
2\frac{\delta L}{\delta h^{\alpha\sigma}}
 h^{\tau\alpha} +
\sqrt{-\gamma}{\psi_{\sigma}^{\ \tau\lambda}}_{;\lambda} =0~.\nonumber
\ena
Using in the first two terms the definitions of 
$\stackrel{m}{t}\!^{\mu\nu}$ and $\stackrel{c}{t}\!^{\mu\nu}$ and formula
(\ref{tuptd}), we arrive at the universal relationship
\bea
-\stackrel{m}{t}\!^{\mu\nu} + \stackrel{c}{t}\!^{\mu\nu}+
{\psi^{\mu\nu \tau}}_{;\tau} -
\frac 2{\sqrt{-\gamma}}\gamma^{\mu\alpha}h^{\nu\beta}\frac{\delta L}
{\delta h^{\alpha\beta}}=0. \label{N5*}
\ena
Assuming that the field equations are satisfied (the last term vanishes)
we can finally conclude that 
\bea
\stackrel{m}{t}\!^{\mu\nu} = ~\stackrel{c}{t}\!^{\mu\nu}  +
{\psi^{\mu\nu \tau}}_{;\tau}~. \nonumber
\ena

	Thus, the metrical and canonical tensors are related by a  
superpotential whose explicit form is given by eq. (\ref{psi0}). 
(This derivation is similar to the one given in \cite{Rosenfeld}.) 
Obviously, the conservation laws are satisfied because 
$
{\psi^{\mu\nu \tau}}_{;\tau ;\nu} \equiv 0~. \nonumber
$

\section{Field Theoretical Formulation of the General Relativity}

	The field theoretical approach to the general relativity treats
gravity as a symmetric tensor field $h^{\mu\nu}$ in Minkowski space-time.
This approach has a long and fruitful history. In fact, in the early
days of special relativity, Poincare and
Einstein himself started from an attempt to give a 
relativistic generalisation
of the Newton law. Even after the acceptance of the geometrical viewpoint,
various aspects of this approach have been worked out in numerous
publications \cite{Papap}--\cite{Deser}, \cite{GPP}, \cite{Gr}
-- to name only a few. (One can also find
references \cite{GZ} useful.)  We will follow a specific 
scheme developed in \cite{GPP} and \cite{Gr}, as a continuation of the line of
reference \cite{Deser}.

	The gravitational field $h^{\mu\nu}(x^{\alpha})$, as well as all 
matter fields, are defined in
the Minkowski space-time with the metric tensor 
$\gamma_{\mu\nu}(x^{\alpha})$:
$ 
d {\sigma}^2 = \gamma_{\mu\nu}dx^{\mu}dx^{\nu}. \nonumber
$ 
The matrix $\gamma^{\mu\nu}$ is the inverse matrix to 
$\gamma_{\mu\nu}$, that is, 
$\gamma^{\alpha\beta}\gamma_{\beta\nu}=\delta^{\alpha}_{\nu}$,
and $\gamma$ is the determinant of the matrix $\gamma_{\mu\nu}$. 
The raising and
lowering of indices are being performed (unless something different is
explicitely stated) with the help of the metric tensor $\gamma_{\mu\nu}$.
The Christoffel symbols associated with $\gamma_{\mu\nu}$ are denoted by
$C^{\tau}_{\ \mu\nu}$, and the covariant derivatives are denoted by a
semicolon ``;''. The curvature tensor of the Minkowski space-time is 
identically
zero:
$
\breve R_{\alpha\beta\mu\nu}(\gamma^{\rho\sigma}) \equiv 0. \nonumber
$

	In terms of classical mechanics, the field variables 
$h^{\mu\nu}$ are the generalised coordinates. Their derivatives
${h^{\mu\nu}}_{;\tau}$ (a third-rank tensor) are the generalised velocities.  
It is also convenient (even if not necessary) to use the generalised momenta
$P^{\alpha}_{\ \mu\nu}$ canonically conjugated 
to the generalised coordinates $h^{\mu\nu}$. The object 
$P^{\alpha}_{\ \mu\nu}$ is a third-rank tensor, symmetric in its
last indices. We will also need a contracted object $P_{\alpha} =
P^{\tau}_{\ \alpha\tau}=\delta^{\nu}_{\mu}P^{\mu}_{\ \alpha\nu}$.

	The use of $h^{\mu\nu}$ and $P^{\tau}_{\ \mu\nu}$ as independent
variables  is an element of the Hamiltonian formalism, which is also known
as the first order variational formalism. We will start from this
presentation, and then will consider the presentation in terms of 
$h^{\mu\nu}$ and ${h^{\mu\nu}}_{;\tau}$. It will be shown that the 
derived field equations
are fully equivalent to the Einstein equations
in the geometrical formulation of the general relativity.

\subsection{Gravitational field equations in terms of generalised
coordinates and momenta}  

	The total action $S$ of the theory consists of the gravitational
part $S^g$ and the matter part $S^m$: $S=S^g + S^m$. We will include the matter 
part in our consideration later on (Sec. VI). The action for the gravitational 
field is
\bea
S^g= \frac 1{c} \int L^g \;d^4x ~, \nonumber
\ena
where the Lagrangian density $L^g$ is
\bea
L^{g}= -\frac{\sqrt{-\gamma}}{2\kappa}\left[{h^{\rho\sigma}}_{;\alpha}
P^{\alpha}_{\ \rho\sigma}
-(\gamma^{\rho\sigma} + h^{\rho\sigma})(P^{\alpha}_{\ \rho\beta}
P^{\beta}_{\ \sigma\alpha} -
\frac1{3}P_{\rho}P_{\sigma})\right]  \label{3}
\ena
and $\kappa = 8\pi G/c^4$. It is now clear that the quantities
$P^{\tau}_{\ \mu\nu}$ are indeed the generalised momenta because 
\bea
-\frac{\sqrt{-\gamma}}{2\kappa} P^{\tau}_{\ \mu\nu}=
\frac{\partial L^g}{\partial {h^{\mu\nu}}_{;\tau}}=
\frac{\partial L^g}{\partial {h^{\mu\nu}}_{,\tau}}  ~. \nonumber
\ena 
The tensor $P^{\tau}_{\mu\nu}$ is related with the tensor $K^{\tau}_{\ \mu\nu}$
originally used in \cite{GPP} by
$$
P^{\tau}_{\ \mu\nu} = -K^{\tau}_{\ \mu\nu} + \frac 1{2} \delta^{\tau}_{\mu}
K_{\nu} + \frac 1{2} \delta^{\tau}_{\nu} K_{\mu}~.
$$

	To make the part of $L^g$, which is quadratic 
in the momenta $P^{\tau}_{\mu\nu}$, more compact, we will also write the
 Lagrangian in the equivalent form:    
\bea
L^{g}= -\frac{\sqrt{-\gamma}}{2\kappa}  \left[{h^{\rho\sigma}}_{;\alpha}
P^{\alpha}_{\ \rho\sigma}
- \frac 1{2}{\Omega^{\rho\sigma\alpha\beta}}_{\omega\tau}
P^{\tau}_{\ \rho\sigma}P^{\omega}_{\ \alpha\beta}
\right] \label{3cf}
\ena
where
\bea
{\Omega^{\rho\sigma\alpha\beta}}_{\omega\tau} &\equiv& \frac 1{2}
 \left[ (\gamma^{\rho\alpha} + h^{\rho\alpha})(\delta^{\sigma}_{\omega}
 \delta^{\beta}_{\tau} -\frac 1{3} \delta^{\sigma}_{\tau}
\delta^{\beta}_{\omega})  +
(\gamma^{\sigma\alpha} + h^{\sigma\alpha})(\delta^{\rho}_{\omega}
 \delta^{\beta}_{\tau} -\frac 1{3} \delta^{\rho}_{\tau}
\delta^{\beta}_{\omega}) +
\right.\nonumber \\
& &\left.
(\gamma^{\rho\beta} + h^{\rho\beta})(\delta^{\sigma}_{\omega}
 \delta^{\alpha}_{\tau} -\frac 1{3} \delta^{\sigma}_{\tau}
\delta^{\alpha}_{\omega}) + 
(\gamma^{\sigma\beta} + h^{\sigma\beta})(\delta^{\rho}_{\omega}
 \delta^{\alpha}_{\tau} -\frac 1{3} \delta^{\rho}_{\tau}
\delta^{\alpha}_{\omega})\right] \label{o}
\ena
and
$
{\Omega^{\mu\nu\alpha\beta}}_{\omega\tau}=
{\Omega^{\nu\mu\alpha\beta}}_{\omega\tau}=
{\Omega^{\mu\nu\beta\alpha}}_{\omega\tau}=
{\Omega^{\alpha\beta\mu\nu}}_{\tau\omega}~. \nonumber
$

	The gravitational field equations are derived by applying 
the variational principle to (\ref{3}) and considering 
the variables $h^{\mu\nu}$ and $P^{\alpha}_{\ \mu\nu}$ as independent. 
In this framework, the field equations are
\bea
 \frac{\partial L^g}{\partial h^{\mu\nu}} -
\left(\frac{\partial L^g}{\partial {h^{\mu\nu}}_{;\tau}}\right)_{;\tau}=0
\;\;\;{\rm and}\;\;\;
\frac{\partial L^g}{\partial P^{\alpha}_{\ \mu\nu}} -
\left(\frac{\partial L^g}{\partial {P^{\alpha}_{\ 
\mu\nu}}_{;\tau}}\right)_{;\tau}= 0~.\label{64'} 
\ena
Obviously, the term $\frac{\partial L^g}{\partial {P^{\alpha}_{\ \mu\nu}}
_{;\tau}}$ 
in (\ref{64'}) is zero for the Lagrangian (\ref{3}).
Calculating the derivatives directly from (\ref{3}) and introducing
the short-hand notations for the corresponding expressions, one obtains
\bea
-\frac{2\kappa}{\sqrt{-\gamma}} \frac{\delta L^g}{\delta h^{\mu\nu}}&\equiv&
r_{\mu\nu}\equiv - P^{\alpha}_{\ \mu\nu ;\alpha} - P^{\alpha}_{\ \mu\beta}
P^{\beta}_{\ \nu\alpha} + \frac1{3} P_{\mu} P_{\nu}= 0~, \label{8} \\
-\frac{2\kappa}{\sqrt{-\gamma}}\frac{\delta L^g}{\delta P^{\tau}_{\ \mu\nu}}
\equiv
f_{\tau}^{\ \mu\nu} &\equiv& \nonumber
{h^{\mu\nu}}_{;\tau} - (\gamma^{\mu\alpha}
+h^{\mu\alpha})P^{\nu}_{\ \alpha\tau} - (\gamma^{\nu\alpha} + h^{\nu\alpha})
P^{\mu}_{\ \alpha\tau} +\\
& & \frac1{3} \delta^{\nu}_{\tau} (\gamma^{\mu\alpha}
+ h^{\mu\alpha}) P_{\alpha} +
 \frac1{3} \delta^{\mu}_{\tau} (\gamma^{\nu\alpha}
+ h^{\nu\alpha}) P_{\alpha} =0~. \label{9}
\ena
Using the $\Omega$-matrix introduced above we can rewrite eq. (\ref{9})
in the compact form:
\bea
 {h^{\mu\nu}}_{;\tau} =
 {\Omega^{\mu\nu\alpha\beta}}_{\omega\tau}P^{\omega}_{\ \alpha\beta}~.
\label{9'}
\ena
Equations $r_{\mu\nu}=0$ and $f_{\tau}^{\ \mu\nu}=0$ form a complete set of
equations in the framework of the first order variational formalism.

\subsection{Field equations in terms of generalised coordinates
and velocities}

	We will need the field equations in terms of the
gravitational field variables  $h^{\mu\nu}$ and their derivatives.
We will derive the equations from the Lagrangian (\ref{3cf}) written
in the form containing the generalised coordinates and velocities.
This is an element of the Lagrangian formalism, known also 
as the second order variational formalism. To implement this program 
one has to consider $P^{\alpha}_{\ \mu\nu}$ as known functions 
of $h^{\mu\nu}$ and $h^{\mu\nu}\ _{;\alpha}$ 
and to use them in the Lagrangian (\ref{3cf}).  

	The link between $h^{\mu\nu}$ and $P^{\tau}_{\ \mu\nu}$ is 
provided by eq. (\ref{9'}). To solve equations (\ref{9'}) 
with respect to $P^{\tau}_{\ \mu\nu}$ we introduce the
matrix ${\Omega^{-1}_{\rho\sigma\mu\nu}}^{\tau\psi}$, which is the
inverse matrix to ${\Omega^{\alpha\beta\mu\nu}}_{\tau\omega}$ and
satisfies the equation  
\bea
{\Omega^{\mu\nu\alpha\beta}}_{\omega\tau}
{\Omega^{-1}_{\rho\sigma\mu\nu}}^{\tau\psi} = \frac 1{2}
\delta^{\psi}_{\omega}(\delta^{\alpha}_{\rho}\delta^{\beta}_{\sigma}+
\delta^{\alpha}_{\sigma}\delta^{\beta}_{\rho})~.\label{o-1} 
\ena
The explicit form of the $\Omega^{-1}$-matrix is not needed for the time
being, but it will be given below where required. We will only use
the symmetry properties of 
the ${\Omega^{-1}_{\mu\nu\rho\sigma}}^{\tau\omega}$ which are 
the same as the symmetry properties of the $\Omega$-matrix:
$
 {\Omega^{-1}_{\mu\nu\rho\sigma}}^{\tau\omega}=
 {\Omega^{-1}_{\nu\mu\rho\sigma}}^{\tau\omega}=
 {\Omega^{-1}_{\mu\nu\sigma\rho}}^{\tau\omega}=
 {\Omega^{-1}_{\rho\sigma\mu\nu}}^{\omega\tau}~. \nonumber
$
By multiplying the both sides of eq. (\ref{9'}) with 
${\Omega^{-1}_{\rho\sigma\mu\nu}}^{\tau\psi}$
one obtains  
\bea
P^{\tau}_{\ \mu\nu} = {\Omega^{-1}_{\rho\sigma\mu\nu}}^{\tau\omega}
{h^{\rho\sigma}}_{;\omega}~. \label{17cf}
\ena

	Now we substitute eq. (\ref{17cf}) into eq. (\ref{3cf}). The Lagrangian
takes the elegant form  
\bea
L^{g}=-\frac{\sqrt{-\gamma}}{4\kappa} 
{\Omega^{-1}_{\rho\sigma\alpha\beta}}^{\omega\tau}
{h^{\rho\sigma}}_{;\tau} {h^{\alpha\beta}}_{;\omega}~ ,
\label{3nso}
\ena
which is manifestly quadratic in the generalised 
velocities  ${h^{\mu\nu}}_{;\tau}$. The dependence on the generalised
coordinates $h^{\mu\nu}$ (as well as on the metric tensor $\gamma_{\mu\nu}$)
is contained in the $\Omega^{-1}$ tensor. The Lagrangian $L^g$
belongs to the class of Lagrangians (\ref{L*}) studied in Sec. II.  

      The field equations in the framework of the second order 
variational formalism are
\bea
\frac{\delta L^g}{\delta h^{\mu\nu}}= \frac{\partial L^g}{\partial h^{\mu\nu}} -
\left(\frac{\partial L^g}{\partial {h^{\mu\nu}}_{;\tau}}\right)_{;\tau} = 0~.
\nonumber
\ena
In more detail, we have
\bea
\frac{\partial{\Omega^{-1}_{\rho\sigma\alpha\beta}}^{\omega\tau}}
{\partial h^{\mu\nu}} {h^{\rho\sigma}}_{;\tau} {h^{\alpha\beta}}_{;\omega}
- 2\left({\Omega^{-1}_{\mu\nu\alpha\beta}}^{\omega\tau}
 {h^{\alpha\beta}}_{;\omega}\right)_{;\tau} =0~. \nonumber
\ena
The first term can be calculated by differentiating (\ref{o-1}) with respect
to $h^{\mu\nu}$ and taking into account (\ref{o}). This gives
\bea
\frac{\partial {\Omega^{-1}_{\rho\sigma\alpha\beta}}^{\omega\tau}}
{\partial h^{\mu\nu}} =- \left(\delta^{\phi}_{\pi}\delta^{\psi}_{\epsilon} -
\frac 1{3}\delta^{\phi}_{\epsilon}\delta^{\psi}_{\pi}\right)
\left[{\Omega^{-1}_{\mu\psi\rho\sigma}}^{\tau\pi}
{\Omega^{-1}_{\nu\phi\alpha\beta}}^{\omega\epsilon} +
{\Omega^{-1}_{\nu\psi\rho\sigma}}^{\tau\pi}
{\Omega^{-1}_{\mu\phi\alpha\beta}}^{\omega\epsilon}\right]~. \nonumber
\ena
The second term requires to recall the rules of the covariant 
differentiation applied to  
the $\Omega^{-1}$ tensor which, in turn, is a function of $h^{\mu\nu}$ and
$\gamma^{\mu\nu}$:  
\bea
{{\Omega^{-1}_{\mu\nu\alpha\beta}}^{\omega\tau}}_{;\tau}=
\frac{\partial {\Omega^{-1}_{\mu\nu\alpha\beta}}^{\omega\tau}}{\partial 
h^{\rho\sigma}}
{h^{\rho\sigma}}_{;\tau} ~. \nonumber
\ena
Combining all together, one arrives at the field equations which are
manifestly the second-order differential
equations in terms of $h^{\mu \nu}$: 
\bea
{\Omega^{-1}_{\mu\nu\alpha\beta}}^{\omega\tau}{h^{\alpha\beta}}_{;\omega ;\tau}
-  \left(\delta^{\phi}_{\pi}\delta^{\psi}_{\epsilon} -
\frac 1{3}\delta^{\phi}_{\epsilon}\delta^{\psi}_{\pi}\right)
\left[2{\Omega^{-1}_{\rho\psi\mu\nu}}^{\tau\pi}
{\Omega^{-1}_{\sigma\phi\alpha\beta}}^{\omega\epsilon} -
{\Omega^{-1}_{\mu\psi\rho\sigma}}^{\tau\pi}
{\Omega^{-1}_{\nu\phi\alpha\beta}}^{\omega\epsilon}\right]
 {h^{\rho\sigma}}_{;\tau}
{h^{\alpha\beta}}_{;\omega}=0~. \label{8cf}
\ena
Certainly, one arrives at exactly the same equations by substituting
$P^{\tau}_{\ \mu\nu}$ found from eq. (\ref{9'}) (see eq.(\ref{17cf})) 
directly into (\ref{8}).

\subsection{Equivalence of the field theoretical and geometrical 
formulations of the general relativity}

	We will now show that the entire mathematical content of the general
relativity (without matter sources, so far) is covered by the Lagrangian
(\ref{3}), or by its equivalent form (\ref{3nso}). We will demonstrate the
equivalence  directly at the level of the field equations, rather than at
the level of the Lagrangian (\ref{3}) and its Hilbert-Einstein counterpart.
The derived field equations (\ref{8}), (\ref{9}) can be rearranged by identical
transformations into the usual Einstein equations. 

	First, we introduce a new tensor field $g^{\mu\nu}(x^{\alpha})$ 
according to the rule:
\bea
\sqrt{-g}g^{\mu\nu}= \sqrt{-\gamma}(\gamma^{\mu\nu} + h^{\mu\nu}) \label{10}
\ena
where $g=det|g_{\mu\nu}|$ and the tensor $g_{\mu\nu}$ is the inverse 
matrix to the $g^{\mu\nu}$ matrix: 
\bea
g^{\mu\alpha}g_{\nu\alpha}= \delta^{\mu}_{\nu}~. \label{g-1}
\ena
Let us emphasize again that the tensor $g_{\mu\nu}$ is the inverse matrix 
to $g^{\mu\nu}$, and not the tensor $g^{\mu\nu}$ with the lowered indices,  
$
g_{\mu\nu}\ne \gamma_{\mu\alpha}\gamma_{\nu\beta} g^{\alpha\beta}.
$
For the time being, we do not assign any physical interpretation to the
tensor field $g_{\mu\nu}$, we only say that the functions 
$g^{\mu\nu}(x^{\alpha})$ and $g_{\mu\nu}(x^{\alpha})$ are calculable
from the functions  
$h^{\mu\nu}(x^{\alpha})$ and $\gamma^{\mu\nu}(x^{\alpha})$ 
according to the given rules (\ref{10}), (\ref{g-1}).

	The introduced quantities allow us to write the  
$\Omega$-matrix as 
\bea
{\Omega^{\rho\sigma\alpha\beta}}_{\omega\tau} &=&
\frac {\sqrt{-g}}{2\sqrt{-\gamma}}
 \left[ g^{\rho\alpha} (\delta^{\sigma}_{\omega}
 \delta^{\beta}_{\tau} -\frac 1{3} \delta^{\sigma}_{\tau}
\delta^{\beta}_{\omega})  +
g^{\sigma\alpha}(\delta^{\rho}_{\omega}
 \delta^{\beta}_{\tau} -\frac 1{3} \delta^{\rho}_{\tau}
\delta^{\beta}_{\omega}) +
\right.\nonumber \\
& &\left.
g^{\rho\beta} (\delta^{\sigma}_{\omega}
 \delta^{\alpha}_{\tau} -\frac 1{3} \delta^{\sigma}_{\tau}
\delta^{\alpha}_{\omega}) + 
g^{\sigma\beta}(\delta^{\rho}_{\omega}
 \delta^{\alpha}_{\tau} -\frac 1{3} \delta^{\rho}_{\tau}
\delta^{\alpha}_{\omega})\right]~.\nonumber
\ena
We can also give the explicit form for the
${\Omega^{-1}_{\mu\nu\rho\sigma}}^{\tau\omega}$.
By multiplying the both sides of (\ref{o-1}) with 
$\frac 1{4} \frac{\sqrt{-\gamma}}{\sqrt{-g}}[2 \delta^{\delta}_{\alpha}
(\delta^{\omega}_{\phi}\delta^{\epsilon}_{\lambda} +
\delta^{\omega}_{\lambda}\delta^{\epsilon}_{\phi})g_{\epsilon\beta}-
g^{\delta\omega}(2g_{\alpha\phi}g_{\beta\lambda}-g_{\alpha\beta}
g_{\phi\lambda})]$ 
one obtains the explicit form of the $\Omega^{-1}$-matrix:
\bea
{\Omega^{-1}_{\mu\nu\rho\sigma}}^{\tau\omega}= \frac 1{4} \frac{\sqrt{-\gamma}}
{\sqrt{-g}}[(\delta^{\tau}_{\mu}\delta^{\pi}_{\nu} +
\delta^{\tau}_{\nu}\delta^{\pi}_{\mu})(\delta^{\omega}_{\rho}
\delta^{\lambda}_{\sigma} +
\delta^{\omega}_{\sigma}\delta^{\lambda}_{\rho})g_{\pi\lambda}
-g^{\tau\omega}(g_{\mu\rho}g_{\nu\sigma} + g_{\nu\rho}g_{\mu\sigma}
- g_{\mu\nu}g_{\rho\sigma})]~. \nonumber
\ena

	We now want to calculate the quantity 
$\Gamma^{\tau}_{\ \mu\nu}$ defined by the expression 
\bea
\Gamma^{\tau}_{\ \mu\nu} =\frac 1{2} g^{\tau\lambda}(g_{\lambda\mu ,\nu} + 
g_{\lambda\nu ,\mu} -
g_{\mu\nu ,\lambda})~. \label{Gamma}
\ena
By replacing the partial derivatives with the covariant ones we get 
\bea
\Gamma^{\tau}_{\ \mu\nu} = C^{\tau}_{\ \mu\nu} +
\frac 1{2} g^{\tau\lambda}(g_{\lambda\mu ;\nu} + g_{\lambda\nu ;\mu} -
g_{\mu\nu ;\lambda})~. \label{Gcov}
\ena
Now we want to trade $g_{\mu\nu ;\tau}$ for $(\sqrt{-g}g^{\mu\nu})_{;\tau}$
in order to have quantities easily expressible 
in terms of $\gamma_{\mu\nu}$ and $h^{\alpha\beta}$.
By differentiating (\ref{g-1}) one obtains
\bea
g_{\mu\nu ;\tau}= -g_{\mu\rho}g_{\nu\sigma}{g^{\rho\sigma}}_{;\tau}~. 
\label{gupgd}
\ena
Using the formula for the differentiation of determinants,
we can write 
\bea
{g^{\rho\sigma}}_{;\tau} = \frac 1{\sqrt{-g}}\left[
(\sqrt{-g}g^{\rho\sigma})_{;\tau} -\frac 1{2}g_{\alpha\beta}g^{\rho\sigma}
(\sqrt{-g}g^{\alpha\beta})_{;\tau}\right]~. \label{g-gg}
\ena
Substituting (\ref{gupgd}) and (\ref{g-gg}) in (\ref{Gcov}) we obtain
\bea
\Gamma^{\tau}_{\ \mu\nu} &=& C^{\tau}_{\ \mu\nu}+\frac 1{2\sqrt{-g}}
\left[-\delta^{\tau}_{\sigma} g_{\mu\rho}(\sg g^{\rho\sigma})_{;\nu} -
\delta^{\tau}_{\sigma} g_{\nu\rho}(\sg g^{\rho\sigma})_{;\mu} +
g^{\tau\lambda}g_{\mu\rho}g_{\nu\sigma}(\sg g^{\rho\sigma})_{;\lambda}+
\right. \nonumber \\
& &\left. \frac 1{2} g_{\alpha\beta}(\delta^{\tau}_{\mu}
(\sg g^{\alpha\beta})_{;\nu} +
\delta^{\tau}_{\nu} (\sg g^{\alpha\beta})_{;\mu} -
g^{\tau\lambda}g_{\mu\nu} (\sg g^{\alpha\beta})_{;\lambda})\right]=
\nonumber \\
& & \frac{1}{\sqrt{-\gamma}} \left(
-{\Omega^{-1}_{\mu\nu\rho\sigma}}^{\lambda\tau} + \frac 1{3}\delta^{\tau}_{\mu}
{\Omega^{-1}_{\tau\nu\rho\sigma}}^{\lambda\tau} + \frac 1{3}\delta^{\tau}_{\nu}
{\Omega^{-1}_{\tau\mu\rho\sigma}}^{\lambda\tau}\right)
(\sg g^{\rho\sigma})_{;\lambda}~. \nonumber
\ena
Finally, taking into account $(\sqrt{-g} g^{\rho\sigma})_{;\alpha}=
\sqrt{-\gamma}{h^{\rho\sigma}}_{;\alpha}$ and recalling (\ref{17cf}),
we arrive at  
\bea
\Gamma^{\tau}_{\ \mu\nu} = C^{\tau}_{\ \mu\nu} 
 -P^{\tau}_{\ \mu\nu} + \frac 1{3} \delta^{\tau}_{\mu}
P_{\nu} + \frac 1{3} \delta^{\tau}_{\nu} P_{\mu}~. \label{G}
\ena

	Now we want to use (\ref{G}) and calculate the quantity $R_{\mu\nu}$
defined by the expression
\bea
R_{\mu\nu} = {\Gamma^{\alpha}_{\ \mu\nu}}_{,\alpha} -\frac1{2}
\Gamma^{\alpha}_{\ \mu\alpha,\nu} -\frac1{2} \Gamma^{\alpha}_{\ \nu\alpha,\mu}
+ \Gamma^{\alpha}_{\ \mu\nu}\Gamma^{\beta}_{\ \alpha\beta}-
\Gamma^{\alpha}_{\ \mu\beta}\Gamma^{\beta}_{\ \nu\alpha}~. \label{Ricci}
\ena
The $C^{\tau}_{\ \mu\nu}$ part of $\Gamma^{\tau}_{\ \mu\nu}$ 
produces a series of terms which combine in the Ricci tensor
$\breve R_{\mu\nu}$ of the flat space-time. 
The ordinary derivative of the tensor $P^{\tau}_{\mu\nu}$
plus all the terms containing the product of
$P^{\alpha}_{\mu\beta}$ with $C^{\beta}_{\ \alpha\nu}$ 
combine in the covariant derivative of $P^{\tau}_{\mu\nu}$. All other
terms produce quadratic combinations of 
$P^{\alpha}_{\mu\beta}$. In the result, we arrive at  
\bea
R_{\mu\nu} = \breve R_{\mu\nu}- \left( P^{\alpha}_{\ \mu\nu ;\alpha} + 
P^{\alpha}_{\ \mu\beta}
P^{\beta}_{\ \nu\alpha} -\frac1{3} P_{\mu} P_{\nu}\right)~. \label{R-r}
\ena
Since $\breve R_{\mu\nu} \equiv 0$ we conclude that the field equations
(\ref{8}) are fully equivalent to the equations 
\bea
R_{\mu\nu} = 0~. \label{R=0}
\ena

	The remaining step is the matter of interpretation. We can now
interpret the quantities $g_{\alpha\beta}$ as the metric tensor of the 
curved space-time:
\bea
ds^2 = g_{\mu\nu}dx^{\mu}dx^{\nu}~. \label{ds2}
\ena
Then, the quantities (\ref{Gamma}) are the Christoffel symbols associated with 
this
metric, and the quantities (\ref{Ricci}) are the Ricci tensor of the curved
space-time. Finally, equations (\ref{R=0}) are the Einstein equations
(without matter sources).

\section{The gravitational energy-momentum tensor}

Being armed with the definitions of the energy-momentum tensor
(Section II), as well as with the gravitational Lagrangian and 
field equations (Section III), we are now in the position to derive
the gravitational energy-momentum tensor. We will derive both tensors,
metrical and canonical, and following the general theory of their 
connection, we will find explicitely the superpotential which
relates them. We will show that the requirement that the 
metrical tensor does not contain second derivatives, and the requirement
that the canonical tensor is symmetric, produce one and the same object
which we call the true energy-momentum tensor. This object satisfies
all the 6 demands listed in the Abstract of the paper.

\subsection{The metrical tensor}

The metrical energy-momentum tensor defined by eq. (\ref{mt})
and derived from the Lagrangian density (\ref{3}) has the following form:
\begin{eqnarray}
\kappa\!\!\stackrel{m}{t}\!^{\mu\nu} =
\frac1{2} \gamma^{\mu\nu}{h^{\rho\sigma}}_{;\alpha}
P^{\alpha}_{\ \rho\sigma} + [\gamma^{\mu\rho}\gamma^{\nu\sigma} -
\frac1{2}\gamma^{\mu\nu}(\gamma^{\rho\sigma} + h^{\rho\sigma})]
(P^{\alpha}_{\ \rho\beta}P^{\beta}_{\ \sigma\alpha} -
 \frac1{3} P_{\rho}P_{\sigma}) + Q^{\mu\nu}, \label{tv}
\end{eqnarray}
where
\bea
Q^{\mu\nu}= \frac1{2}(\delta^{\mu}_{\rho}\delta^{\nu}_{\sigma}+
\delta^{\nu}_{\rho}\delta^{\mu}_{\sigma})
[-\gamma^{\rho\alpha}
h^{\beta\sigma}P^{\tau}_{\ \alpha\beta} +(\gamma^{\alpha\tau}h^{\beta\rho} -
\gamma^{\alpha\rho}h^{\beta\tau})P^{\sigma}_{\ \alpha\beta}]_{;\tau}~.
\nonumber
\ena
Expression (\ref{tv}) was obtained by direct calculation of the 
variational derivative (or alternatively, see Appendix B)
\bea
\frac{\delta L}{\delta \gamma_{\mu\nu}} =
\frac{\partial L}{\partial \gamma_{\mu\nu}} -
\left(\frac{\partial L}{\partial \gamma_{\mu\nu ,\tau}}
\right)_{,\tau}, \label{vdg}
\ena
and no further rearrangements have been 
done. Obviously, tensor (\ref{tv}) is symmetric in
its components, but it contains second order 
derivatives of $h^{\mu\nu}$ which enter
the expression through the $Q^{\mu\nu}$ term. We want to single out the
second derivatives of $h^{\mu\nu}$ explicitely.  

By making identical transformations of the $Q^{\mu\nu}$ term 
one can show that the $Q^{\mu\nu}$ contains a term proportional to
$r_{\mu\nu}$ and terms proportional to ${f_{\tau}}^{\mu\nu}$
and its derivatives. All these terms are equal to
zero according to the field equations (\ref{8}) and (\ref{9}).
After removing these terms, the remaining expression for $Q^{\mu\nu}$ 
is as follows
\bea
Q^{\mu\nu}&=& \frac 1{2}(\delta^{\mu}_{\rho}\delta^{\nu}_{\sigma}+
\delta^{\nu}_{\rho}\delta^{\mu}_{\sigma}) \left[
q^{\alpha\beta\rho\sigma}(P^{\pi}_{\ \alpha\lambda}P^{\lambda}_{\ \beta\pi} -
\frac 1{3}P_{\alpha}P_{\beta}) -{q^{\alpha\beta\rho\sigma}}_{;\tau}
P^{\tau}_{\ \alpha\beta} - \right. \nonumber\\
& & \left.
\frac 1{4}({h^{\rho\alpha}}_{;\alpha}
{h^{\sigma\beta}}_{;\beta} - {h^{\rho\alpha}}_{;\beta}{h^{\sigma\beta}}_{;\alpha})
+ \frac 1{2} (h^{\rho\tau}h^{\sigma\lambda} - h^{\tau\lambda}h^{\rho\sigma}
)_{;\lambda ;\tau}\right] \label{Qtr}
\ena
where
\bea
q^{\alpha\beta\rho\sigma}\equiv\frac 1{2}\left[
h^{\sigma\alpha}\gamma^{\rho\beta}+ h^{\rho\alpha}\gamma^{\sigma\beta}+
h^{\sigma\beta}\gamma^{\rho\alpha}+ h^{\rho\beta}\gamma^{\sigma\alpha}+
h^{\sigma\alpha} h^{\rho\beta}+ h^{\rho\alpha}h^{\sigma\beta}
- h^{\rho\sigma}(\gamma^{\alpha\beta}+h^{\alpha\beta}) \right].   \label{q}
\ena
The remaining expression (\ref{Qtr}), together with other terms 
in (\ref{tv}), reduce the $\kappa\!\!\stackrel{m}{t}\!^{\mu\nu}$ to  
\bea
\kappa\!\!\stackrel{m}{t}\!^{\mu\nu}|_r & = &  -\left[(\gamma^{\alpha\nu} + 
h^{\alpha\nu})(\gamma^{\beta\mu} +h^{\beta\mu})
- \frac 1{2}(\gamma^{\alpha\beta} + h^{\alpha\beta})(\gamma^{\mu\nu}+
h^{\mu\nu})\right]_{;\tau} P^{\tau}_{\ \alpha\beta} +
 \nonumber \\
& & \left[(\gamma^{\alpha\nu} + h^{\alpha\nu})(\gamma^{\beta\mu} +h^{\beta\mu})
- \frac 1{2}(\gamma^{\alpha\beta} + h^{\alpha\beta})(\gamma^{\mu\nu}+
h^{\mu\nu})\right]\left(P^{\sigma}_{\ \beta\rho}P^{\rho}_{\ \alpha\sigma} -
\frac 1{3} P_{\alpha}P_{\beta}\right)-
 \nonumber \\
& & \frac 1{2}({h^{\nu\alpha}}_{;\alpha}
{h^{\mu\beta}}_{;\beta} - {h^{\mu\alpha}}_{;\beta}{h^{\nu\beta}}_{;\alpha})
+ \frac1 {4} (-2h^{\mu\nu}h^{\alpha\beta} + h^{\mu\alpha}h^{\nu\beta} +
h^{\nu\alpha}h^{\mu\beta})_{;\alpha ;\beta} 
\label{redmt}
\ena
where the subscript  $|_r$ indicates that the energy-momentum tensor 
was reduced on the equations of motion. 

The last group of terms in (\ref{redmt}) still contains second 
order derivatives 
of $h^{\mu\nu}$, but they all can be removed by a special 
choice of superpotential. Indeed, the symmetric function   
$\Phi^{\mu\nu}$ participating in (\ref{mt'}) 
and satisfying ${\Phi^{\mu\nu}}_{;\nu}\equiv 0$ 
can be written as   
\bea
\Phi^{\mu\nu} = (\phi^{\mu\nu\alpha\beta} +
\phi^{\nu\mu\alpha\beta})_{;\alpha ;\beta} \label{fi-psi}
\ena
where
\bea
\phi^{\mu\nu\alpha\beta} =- \phi^{\alpha\nu\mu\beta}=-\phi^{\mu\beta\alpha\nu}=
\phi^{\nu\mu\beta\alpha}. \label{sympsi}
\ena
To remove all the second order derivatives, we require 
\bea
 \frac1 {4} (-2h^{\mu\nu}h^{\alpha\beta} + h^{\mu\alpha}h^{\nu\beta} +
h^{\nu\alpha}h^{\mu\beta})_{;\alpha ;\beta} + 
(\phi^{\mu\nu\alpha\beta} + \phi^{\nu\mu\alpha\beta})_{;\alpha ;\beta}=0.
\label{mpsi}
\ena
The unique solution to this equation (up to trivial additive terms which
can possibly contain $\gamma^{\mu\nu}$ but not the field
variables $h^{\mu\nu}$) is:   
\bea
\phi^{\mu\nu\alpha\beta}= 
\frac 1{4} (h^{\alpha\beta}h^{\mu\nu} -h^{\alpha\nu}h^{\beta\mu}).
\label{psi}
\ena

With the help of the superpotential (\ref{psi}), we can now cancel out 
the terms 
$ \frac1 {4} (-2h^{\mu\nu}h^{\alpha\beta} + h^{\mu\alpha}h^{\nu\beta} +
h^{\nu\alpha}h^{\mu\beta})_{;\alpha ;\beta}$.
The remaining part of (\ref{redmt}) does not contain any 
second order derivatives at all. To write the remaining part in a
more compact form, we replace the 
generalised momenta by the generalised velocities with the help
of (\ref{17cf}), and use the shorter expressions    
$g_{\alpha\beta}$ and $g^{\alpha\beta}$ according to their
definitions (\ref{10}) and (\ref{g-1}). As a result, the metrical 
energy-momentum tensor (\ref{tv}), transformed with the help of the field 
equations and an allowed superpotential, takes the 
following explicit form: 
\begin{eqnarray}
\kappa t^{\mu\nu} &=&
\frac1{4} [2 {h^{\mu\nu}}_{;\rho} 
{h^{\rho\sigma}}_{;\sigma} 
-2{h^{\mu\alpha}}_{;\alpha}{h^{\nu\beta}}_{;\beta}
+
2g^{\rho\sigma}g_{\alpha\beta} {h^{\nu\beta}}_{;\sigma} 
{h^{\mu\alpha}}_{;\rho} +
 g^{\mu\nu}g_{\alpha\rho} {h^{\alpha\beta}}_{;\sigma} 
{h^{\rho\sigma}}_{;\beta}-
\nonumber \\ 
& & 
  2g^{\mu\alpha}g_{\beta\rho} {h^{\nu\beta}}_{;\sigma}{h^{\rho\sigma}}_
{;\alpha} -
2g^{\nu\alpha} g_{\beta\rho} {h^{\mu\beta}}_{;\sigma}
{h^{\rho\sigma}}_{;\alpha} +
\nonumber \\
& &  \frac1{4}(2g^{\mu\delta}g^{\nu\omega} -g^{\mu\nu}g^{\omega\delta})
(2g_{\rho\alpha}g_{\sigma\beta} - 
 g_{\alpha\beta}g_{\rho\sigma})
{h^{\rho\sigma}}_{;\delta}
{h^{\alpha\beta}}_{;\omega}]\label{truemt}
\end{eqnarray}
where $g_{\alpha\beta}$ and $g^{\alpha\beta}$ are short-hand notations for
the quantities (\ref{10}), (\ref{g-1}).
This object is a tensor with respect to arbitrary coordinate transformations,
symmetric in its components, conserved due to the field equations, free of
second derivatives of $h^{\mu\nu}$, and unique up to additive terms not
containing $h^{\mu\nu}$. This derivation required the use of an allowed
superpotential. The last step is to show that the 
energy-momentum tensor (\ref{truemt}) can also be derived according to the
original definition (\ref{mt}), without resorting to the use of a
superpotential. The tensor (\ref{truemt}) will be derived from
a modified Lagrangian, which produces exactly the same field equations as 
(\ref{8}) and (\ref{9}). This is what we will do now.

\subsection{The constrained variational principle}

Let us write the modified Lagrangian in the form
\begin{eqnarray}                             
L^{g} = 
-\frac{\sqrt{-\gamma}}{2\kappa}\left[{h^{\rho\sigma}}_{;\alpha}P^{\alpha}_{\ 
\rho\sigma}
-(\gamma^{\rho\sigma} + h^{\rho\sigma})(P^{\alpha}_{\ \rho\beta}P^{\beta}_{\ 
\sigma\alpha} -
\frac1{3}P_{\rho}P_{\sigma}) + 
 \Lambda^{\alpha\beta\rho\sigma} \breve R_{\alpha\rho\beta\sigma}\right]
 \label{l'}
\end{eqnarray}
where $\breve R_{\alpha\rho\beta\sigma}$ is the curvature tensor 
constructed from $\gamma_{\mu\nu}$. Obviously, we have added zero to 
the original Lagrangian,
but this is a typical  way of incorporating  a constraint (in our case,
$\breve R_{\alpha\rho\beta\sigma}=0$) by means of the undetermined
Lagrange multipliers. The infinitesimal variation (\ref{lim}) of the metric
tensor $\gamma_{\mu\nu}$ 
(and even its exponentiated finite version) do not change the condition
$\breve R_{\alpha\rho\beta\sigma}=0$. 
The multipliers $\Lambda^{\alpha\beta\rho\sigma}$ form a tensor which depends
on $\gamma^{\mu\nu}$ and $h^{\mu\nu}$ and satisfy
\bea
\Lambda^{\alpha\beta\rho\sigma}=
-\Lambda^{\rho\beta\alpha\sigma}=
-\Lambda^{\alpha\sigma\rho\beta}=
\Lambda^{\beta\alpha\sigma\rho}. \label{symlam}
\ena
The variational derivative of 
$\Lambda^{\alpha\beta\rho\sigma}\breve R_{\alpha\rho\beta\sigma}$
with respect to the metric tensor 
$\gamma_{\mu\nu}$ is not zero, and therefore the added term will 
affect the metrical energy-momentum tensor. However, the added term does
not change the field equations, since the variational derivative of this
term with respect
to the field variables $h^{\mu\nu}$ will be multiplied by the 
$\breve R_{\alpha\rho\beta\sigma}$ and hence will vanish due to the
constraint. 
 
The metrical energy-momentum tensor (\ref{mt}) directly derived 
from (\ref{l'}) is now modified as compared with (\ref{tv}):
\begin{eqnarray}
\kappa\!\!\stackrel{m}{t}\!^{\mu\nu}|_c &=&
\frac1{2} \gamma^{\mu\nu}{h^{\rho\sigma}}_{;\alpha}
P^{\alpha}_{\ \rho\sigma} + [\gamma^{\mu\rho}\gamma^{\nu\sigma} -
\frac1{2}\gamma^{\mu\nu}(\gamma^{\rho\sigma} + h^{\rho\sigma})]
(P^{\alpha}_{\ \rho\beta}P^{\beta}_{\ \sigma\alpha} -
\frac1{3} P_{\rho}P_{\sigma}) +\nonumber \\
& &   Q^{\mu\nu}
-(\Lambda^{\mu\nu\alpha\beta} +
\Lambda^{\nu\mu\alpha\beta})_{;\alpha ;\beta} \label{tv'}
\end{eqnarray}
where the subscript $|_c$ indicates that the Lagrangian (\ref{l'})
has been used. 
The entire modification amounts to the last two terms 
(with double derivatives) in (\ref{tv'}), which immediately suggests
its connection to modifications at the expense of superpotentials
(\ref{fi-psi}), (\ref{sympsi}). 
(For a detailed derivation of the last two terms see Appendix B.) 
As before, the tensor 
$\kappa\!\!\stackrel{m}{t}\!\!^{\mu\nu}|_c$ contains second derivatives
of $h^{\mu\nu}$ in the $Q^{\mu\nu}$ term. But the originally undetermined 
multipliers $\Lambda^{\alpha\beta\rho\sigma}$ will now be determined.  
They can be chosen in such a way that the remaining 
second derivatives of $h^{\mu\nu}$ (which could not be
excluded at the field equations) can now be removed. The equations to be
solved are similar to equations (\ref{mpsi}). Their unique solution is  
\bea
\Lambda^{\mu\nu\alpha\beta}= 
-\frac 1{4} (h^{\alpha\beta}h^{\mu\nu} -h^{\alpha\nu}h^{\beta\mu}). \nonumber
\ena
Thus, the energy-momentum tensor (\ref{truemt}) satisfies the last remaining  
demand: it can be derived in a regular prescribed way (\ref{mt}) from the
Lagrangian (\ref{l'}).

\subsection{The canonical tensor}

The gravitational energy-momentum tensor (\ref{truemt}) satisfying all the
necessary demands has been derived along the ``metrical route''. We will
now show that the symmetrisation procedure of the canonical tensor leads
to the same object (\ref{truemt}).  

The canonical energy-momentum tensor (\ref{cct}) directly calculated 
from the Lagrangian density (\ref{3nso}) has the form
\bea
\kappa\!\stackrel{c}{t}\!^{\mu\nu} = - \frac 1{4} (2 \gamma^{\nu\omega}
{\Omega^{-1}_{\rho\sigma\alpha\beta}}^{\mu\tau} - \gamma^{\mu\nu}
{\Omega^{-1}_{\rho\sigma\alpha\beta}}^{\omega\tau}){h^{\rho\sigma}}_{;\tau}
{h^{\alpha\beta}}_{;\omega}. \nonumber 
\ena  
It is convenient to use here and below the quantity 
$P^{\tau}_{~\mu\nu}$ as a short-hand notation for
${\Omega^{-1}_{\mu\nu\alpha\beta}}^{\omega\tau}{h^{\alpha\beta}}_{;\omega}$
in agreement with (\ref{17cf}).
Then, the $\kappa \stackrel{c}{t}\!^{\mu\nu}$
takes the compact form  
\bea
\kappa\!\stackrel{c}{t}\!^{\mu\nu} = -\frac 1{2}\gamma^{\mu\tau}
P^{\nu}_{\ \alpha\beta} {h^{\alpha\beta}}_{;\tau}
+ \frac 1{4}\gamma^{\mu\nu}P^{\tau}_{\ \alpha\beta} {h^{\alpha\beta}}_{;\tau}.
\label{ct}
\ena
As expected, the canonical tensor $\stackrel{c}{t}\!^{\mu\nu}$ 
is not symmetric. It can be made symmetric (see II~A) by an appropriate
choice of $\Psi^{\mu\nu}$. We will do this on the basis of the universal 
relationship (\ref{N5*}) between the (symmetric) 
$\stackrel{m}{t}\!^{\mu\nu}$ and the (non-symmetric)  
$\stackrel{c}{t}\!^{\mu\nu}$. 

The relationship in question is 
\bea
-\kappa\!\!\stackrel{m}{t}\!^{\mu\nu} + \kappa\!\stackrel{c}{t}\!^{\mu\nu} +~
\kappa {\psi^{\mu\nu\tau}}_{;\tau} +
\gamma^{\mu\alpha}h^{\nu\beta}r_{\alpha\beta}=0 \label{S1}
\ena
where $\psi^{\mu\nu\tau}$ is calculated from the 
Lagrangian (\ref{3nso}) according to (\ref{psi0}): 
\bea
\kappa \psi^{\mu\nu\tau} = \frac 1{2} [P^{\tau}_{\alpha\beta}
(\gamma^{\mu\alpha} h^{\nu\beta} -\gamma^{\alpha\nu}h^{\mu\beta}) +
P^{\mu}_{\alpha\beta}
(\gamma^{\alpha\tau} h^{\nu\beta} -\gamma^{\alpha\nu}h^{\tau\beta})  +
P^{\nu}_{\alpha\beta}
(\gamma^{\alpha\tau} h^{\mu\beta} -\gamma^{\alpha\mu}h^{\tau\beta})]. \label{S2}
\ena
As any superpotential does, the tensor $\psi^{\mu\nu\tau}$ 
satisfies the requirement 
${\psi^{\mu\nu\tau}}_{;\tau;\nu} \equiv 0$. 
One can easily check the validity of (\ref{S1}) if one combines (\ref{tv}),
(\ref{ct}),
(\ref{S2}), (\ref{8}) and uses the following identities:
\bea
(\gamma^{\alpha\beta} +h^{\alpha\beta})\left(P^{\rho}_{\ \alpha\sigma}
P^{\sigma}_{\ \beta\rho} -\frac 1{3} P_{\alpha}P_{\beta}\right)\equiv
\frac 1{2} {h^{\rho\sigma}}_{;\alpha}P^{\alpha}_{\ \rho\sigma}~, \nonumber\\
(\gamma^{\nu\beta} +h^{\nu\beta})\left(P^{\rho}_{\ \alpha\sigma}
P^{\sigma}_{\ \beta\rho} -\frac 1{3} P_{\alpha}P_{\beta}\right)\equiv
 {h^{\nu\sigma}}_{;\rho}P^{\rho}_{\ \alpha\sigma} -
 \frac 1{2} {h^{\rho\sigma}}_{;\alpha}P^{\nu}_{\ \rho\sigma}~.  \nonumber
\ena

We now assume that the field equations are satisfied. The last term in
(\ref{S1}) drops out. The metrical tensor $\stackrel{m}{t}\!\!^{\mu\nu}$ 
reduced at the equations of motion is given by (\ref{redmt}). We need also
to reduce the third term in (\ref{S1}) at the equations of motion. First, we
differentiate the expression (\ref{S2}) and make identical transformations to 
rearrange the ${\psi^{\mu\nu\tau}}_{;\tau}$: 
\bea
\kappa{\psi^{\mu\nu\tau}}_{;\tau} &=& \left[h^{\mu\beta}(\gamma^{\nu\alpha}+
h^{\nu\alpha}) -
\frac 1{2}h^{\mu\nu}(\gamma^{\alpha\beta}+ h^{\alpha\beta})\right]
r_{\alpha\beta}
 +\chi^{\mu\nu} + \nonumber \\
& & \frac1 {4} (-2h^{\mu\nu}h^{\alpha\beta} + h^{\mu\alpha}h^{\nu\beta} +
h^{\nu\alpha}h^{\mu\beta})_{;\alpha ;\beta}~, \label{psi-hi}
\ena
where 
\bea
\chi^{\mu\nu} &=&
-\left[h^{\mu\beta}(\gamma^{\nu\alpha} + h^{\nu\alpha})-
\frac 1{2} h^{\mu\nu}(\gamma^{\alpha\beta}+ h^{\alpha\beta})\right]_{;\tau}
P^{\tau}_{\ \alpha\beta} + \nonumber \\
& & \left[h^{\mu\beta}(\gamma^{\nu\alpha} + h^{\nu\alpha})-
\frac 1{2} h^{\mu\nu}(\gamma^{\alpha\beta}+ h^{\alpha\beta})\right]_{;\tau}
\left(P^{\rho}_{\ \alpha\sigma}
P^{\sigma}_{\ \beta\rho} -\frac 1{3} P_{\alpha}P_{\beta}\right)- \nonumber \\
& &\frac 1{2}({h^{\nu\alpha}}_{;\alpha}
{h^{\mu\beta}}_{;\beta} - {h^{\mu\alpha}}_{;\beta}{h^{\nu\beta}}_{;\alpha})~.
\nonumber
\ena
Now we drop the term proportional
to the field equations. The remaining part of 
${\psi^{\mu\nu\tau}}_{;\tau}$ is
\bea
\kappa {\psi^{\mu\nu\tau}}_{;\tau}|_r &=&
\chi^{\mu\nu} +
\frac1 {4} (-2h^{\mu\nu}h^{\alpha\beta} + h^{\mu\alpha}h^{\nu\beta} +
h^{\nu\alpha}h^{\mu\beta})_{;\alpha ;\beta}~. \label{S4}
\ena

On the field equations, the original relationship (\ref{S1}) reduces to 
\bea
 \stackrel{m}{t}\!^{\mu\nu}|_r = \stackrel{c}{t}\!^{\mu\nu} +
~{\psi^{\mu\nu\tau}}_{;\tau}|_r .\label{S5'}
\ena
The last term in expression (\ref{S4}) for  
${\psi^{\mu\nu\tau}}_{;\tau}|_r$ cancels out with exactly the same term
in expression (\ref{redmt}) for  
$\stackrel{m}{t}\!\!^{\mu\nu}|_r$. After this cancellation, what is left
on the left hand side of (\ref{S5'}) is the metrical tensor   
$t^{\mu\nu}$ described by formula (\ref{truemt}). On the right
hand side of (\ref{S5'}) we will get a symmetrised (subscript $|_s$) canonical 
tensor 
$\kappa\!\stackrel{c}{t}\!^{\mu\nu}|_s =~\kappa\!\stackrel{c}{t}\!^{\mu\nu}
+ \chi^{\mu\nu}$.
Note, that since ${\psi^{\mu\nu\tau}}_{;\tau;\nu} \equiv 0$, it follows 
from (\ref{psi-hi}) that  
${\chi^{\mu\nu}}_{;\nu} = 0$ on the field equations $r_{\alpha\beta}=0$.
Thus, we arrive at the equality 
\bea
t^{\mu\nu}=~\stackrel{c}{t}\!^{\mu\nu}|_s~. \nonumber
\ena
Since the metrical tensor (\ref{truemt}) satisfies all the 6 demands listed
above, and since it can also be obtained as a result of symmetrisation of the
canonical tensor, we call it true energy-momentum tensor (and write it
without any labels or subscripts).

 \section{Gravitational field equations with gravitational 
energy-momentum tensor}

	We have derived the gravitational (true) energy-momentum tensor
(\ref{truemt}) from the gravitational Lagrangian $L^g$ according to the 
general definition (\ref{mt}). We know that the conservation 
laws ${t^{\mu\nu}}_{;v} = 0$ are guaranteed on solutions to the field 
equations. The non-linear nature of the gravitational 
field $h^{\mu\nu}$ makes the field
a source for itself. The question arises as for how the $\kappa t^{\mu\nu}$
participates in the field equations. To answer this question one needs
to rearrange the field equations and single out  
the $\kappa t^{\mu\nu}$ explicitely. One can proceed either from 
equations (\ref{8}) or from equations (\ref{8cf}). A simpler way is to 
take the 
following linear combination of the field equations (\ref{8}):
\bea
\left[(\gamma^{\alpha\nu} + 
h^{\alpha\nu})(\gamma^{\beta\mu} +h^{\beta\mu})
- \frac 1{2}(\gamma^{\alpha\beta} + h^{\alpha\beta})(\gamma^{\mu\nu}+
h^{\mu\nu})\right] r_{\alpha\beta} =0 \label{fe-tem1}
\ena
and to use the link (\ref{17cf}) in order to exclude $P^{\alpha}_{\ \mu\nu}$.
After putting all the terms in a necessary order, the field equations
(\ref{fe-tem1}) take the following form
\bea
\frac{1}{2}[(\gamma^{\mu\nu}+ h^{\mu\nu})(\gamma^{\alpha\beta}+ 
h^{\alpha\beta})- (\gamma^{\mu\alpha}+h^{\mu\alpha})
(\gamma^{\nu\beta}+h^{\nu\beta})]_{;\alpha;\beta} = 
\frac1{4} [2 {h^{\mu\nu}}_{;\rho} 
{h^{\rho\sigma}}_{;\sigma} 
-  \nonumber \\
2{h^{\mu\alpha}}_{;\alpha}{h^{\nu\beta}}_{;\beta}
+
2g^{\rho\sigma}g_{\alpha\beta} {h^{\nu\beta}}_{;\sigma} 
{h^{\mu\alpha}}_{;\rho} 
+g^{\mu\nu}g_{\alpha\rho} {h^{\alpha\beta}}_{;\sigma} 
{h^{\rho\sigma}}_{;\beta} 
- 2g^{\mu\alpha}g_{\beta\rho} {h^{\nu\beta}}_{;\sigma}{h^{\rho\sigma}}_
{;\alpha} -  \nonumber \\
 2g^{\nu\alpha} g_{\beta\rho} {h^{\mu\beta}}_{;\sigma}
{h^{\rho\sigma}}_{;\alpha} +
\frac1{4}(2g^{\mu\delta}g^{\nu\omega} -g^{\mu\nu}g^{\omega\delta})
(2g_{\rho\alpha}g_{\sigma\beta}- g_{\alpha\beta}g_{\rho\sigma})
{h^{\rho\sigma}}_{;\delta}
{h^{\alpha\beta}}_{;\omega}]. \label{56}
\ena
On the right-hand side of (\ref{56}) we have exactly the energy-momentum 
tensor (\ref{truemt}), so the field equations can be written 
\bea
\frac{1}{2}[(\gamma^{\mu\nu}+ h^{\mu\nu})(\gamma^{\alpha\beta}+ 
h^{\alpha\beta})- (\gamma^{\mu\alpha}+h^{\mu\alpha})
(\gamma^{\nu\beta}+h^{\nu\beta})]_{;\alpha;\beta} = 
\kappa t^{\mu\nu}. 
\label{55} 
\ena
The left-hand side of this equation is the generalised differential
wave (d'Alembert) operator. So, the gravitational energy-momentum
tensor $\kappa t^{\mu\nu}$ is not, and should not be, a source 
term in the ``right-hand side of the Einstein equations", but it is a
source term for the generalised wave operator.

Replacing the sum 
$(\gamma^{\mu\nu}+ h^{\mu\nu})$ by the shorter expression $g^{\mu\nu}$
according to the definition (\ref{10}), we can also write the gravitational
field equations (\ref{55}) in the form  
\bea
\frac1{2} [(-g)(g^{\mu\nu} g^{\alpha\beta} - g^{\mu\alpha}
g^{\nu\beta}) ]_{;\alpha ;\beta} = \kappa t^{\mu\nu}~. \label{gg-t}
\ena
We know that the field equations (\ref{gg-t}) are fully equivalent to the
Einstein equations (see Sec III). In the geometrical approach to the
general relativity one interprets the quantities
$g_{\mu\nu}$ as the metric tensor of a curved space-time (\ref{ds2}).
It is interesting to ask if there exists an object in the geometrical
approach, which would be somehow related to the energy-momentum tensor   
$t^{\mu\nu}$ derived here.
[The description of energy in the general relativity is, of course,
a matter of a long time effort by many people who used different
approaches. We would like to
mention at least some of works \cite{ADM}--\cite{MagSok},
 which influenced our understanding of
the problem.]  
For this purpose we use for the first
time the available coordinate freedom and introduce the Lorentzian
coordinates. This means that the metric tensor  
$\gamma_{\mu\nu}(x^{\alpha})$ is being transformed by a coordinate
transformation to the usual
constant matrix $\eta_{\mu\nu}$. In other words, one makes   
$\gamma_{00}=1, \gamma_{11}=\gamma_{22}=\gamma_{33}=-1$, the rest
of components zeros, and the determinant $\gamma = -1$. Then, all 
the covariant derivatives can be replaced by the ordinary ones, and
all the derivatives of the metric tensor $\gamma_{\mu\nu}$ vanish. 
Writing the expression (\ref{truemt}) for $t^{\mu\nu}$ in Lorentzian 
coordinates (subscript $|_L$) and using quantities $g^{\mu\nu}$
instead of $h^{\mu\nu}$ one finds that 
\bea
t^{\mu\nu}|_L = (-g) t^{\mu\nu}_{LL} \nonumber
\ena
where $t^{\mu\nu}_{LL}$ is the Landau-Lifshitz pseudotensor \cite{LL}.
The field equations (\ref{gg-t}) written in Lorentzian coordinates take 
the form  
\bea
\frac1{2} [(-g)(g^{\mu\nu} g^{\alpha\beta} - g^{\mu\alpha}
g^{\nu\beta}) ]_{,\alpha, \beta}= \kappa (-g) t^{\mu\nu}_{LL}. \nonumber
\ena
So, the object most closely related to the derived energy-momentum
tensor $t^{\mu\nu}$ is the Landau-Lifshitz pseudotensor  
$t^{\mu\nu}_{LL}$ times $(-g)$. Their numerical values (but not the
transformation properties, unless for linear coordinate
transformations) are the same at least under some conditions.

\section{Gravitational field with matter sources}

We will now include in our consideration matter fields interacting 
with the gravitational field. One or several matter fields are denoted by 
$\phi_A$, where $A$ is some general index.

\subsection{Gravitational field equations and the energy-momentum tensor
for the matter fields}

The total action in presence of the matter sources is
\bea
S=\frac 1{c} \int (L^g + L^m)\; d^4x  \label{58}
\ena
where $L^g$ is the gravitational Lagrangian (\ref{3}) and $L^m$ is
the matter
Lagrangian which includes interaction of the matter fields with the 
gravitational field.
We assume the universal coupling of the gravitational field to all other
physical fields, that is, we assume that the $L^m$ depends on the 
gravitational field variables $h^{\mu\nu}$ in a specific manner:  
\bea
L^m=L^m\left[\sqrt{-\gamma} (\gamma^{\mu\nu} + h^{\mu\nu}); 
(\sqrt{-\gamma} (\gamma^{\mu\nu} + h^{\mu\nu}))_{,\alpha}; 
\phi_A; \phi_{A,\alpha}\right].
\label{coupl}
\ena
It was shown \cite{GPP} that the $L^m$ must depend on  
$h^{\mu\nu}$ and $\gamma^{\mu\nu}$ only through the combination 
$\sqrt{-\gamma} (\gamma^{\mu\nu} + h^{\mu\nu})$, if one wants 
the matter energy-momentum tensor $ \tau^{\mu\nu}$
to participate in the gravitational field equations
at the equal footing with the gravitational energy-momentum 
tensor, that is, through the total energy-momentum tensor which is 
the sum of the two. 
The matter energy-momentum tensor $ \tau^{\mu\nu}$
is defined by the previously
discussed (see Sec. II) universal formula 
\bea
\tau^{\mu\nu}= -\frac2{\sqrt{-\gamma}} \frac{\delta L^m}
{\delta \gamma_{\mu\nu}} .
\ena

Let us now turn to the derivation of the field equations. The gravitational
equations are derived by applying the variational principle to the 
gravitational variables in the total Lagrangian. The previously derived
equations (\ref{9}) remain unchanged since we assume (for simplicity) that
the $L^m$ does not contain $P^{\alpha}_{\ \mu\nu}$. However, equations
(\ref{8}) are changed and take now the form   
\bea
r_{\mu\nu} -\frac{2\kappa}{\sqrt{-\gamma}}
 \frac{\delta L^m}{\delta h^{\mu\nu}} = 0. \label{60}
\ena
As for the matter field equations, they are derived by applying the 
variational principle to the matter variables in the total Lagrangian,
which means:
$
\frac{\delta L^m}{\delta \phi_A}=0. \nonumber
$
The concrete form of the matter field equations will not be needed. 

We know (Sec. V A) that equations (\ref{60}) without the term caused by
$L^m$ are equivalent to equations (\ref{55}), where 
$\kappa t^{\mu\nu}$ is given by formula (\ref{truemt}). 
We want to show that the source term in the right-hand side of the
gravitational equations becomes now, in presence of $L^m$, 
the total energy-momentum tensor.

Let us start from the contribution provided by $L^g$.
Since the procedure of reduction of 
$\kappa\!\!\stackrel{m}{t}\!^{\mu\nu}$ to the final form 
$\kappa t^{\mu\nu}$ (\ref{truemt}) 
involved the use of the equations of motion (\ref{8}), which
are now modified to (\ref{60}), the gravitational part of the total 
energy-momentum tensor will also be modified, as compared with
(\ref{truemt}).     
Using (\ref{60}) instead of (\ref{8}), and getting rid of the second
derivatives of $h^{\mu\nu}$ in the same way as before, one obtains
\bea
\kappa t^{\mu\nu}|_m = \kappa t^{\mu\nu} + q^{\alpha\beta\mu\nu}
\frac{2\kappa}{\sqrt{-\gamma}}\frac{\delta L^m}{\delta h^{\alpha\beta}} 
\nonumber
\ena
where the subscript $|_m$ indicates that the derivation has been done in
presence of the matter fields. The $\kappa t^{\mu\nu}$ is given of course
by the same formula (\ref{truemt}), and quantities $q^{\alpha\beta\mu\nu}$
are given by formula (\ref{q}).  Let us now turn to $\tau^{\mu\nu}$.  
The universal coupling in the Lagrangian (\ref{coupl}), that is, the
fact that $h^{\mu\nu}$ and $\gamma^{\mu\nu}$  enter the $L^m$ only in the
combination $\sqrt{-\gamma} (\gamma^{\mu\nu} + h^{\mu\nu})$, 
allows us to relate
$\frac{\delta L^m}{\delta \gamma_{\mu\nu}}$ with 
$\frac{\delta L^m}{\delta h^{\mu\nu}}$. After necessary transformations,
one obtains  
\bea
\tau^{\mu\nu} = \left[ 2\gamma^{\mu\rho}\gamma^{\nu\sigma} -
\gamma^{\mu\nu}( \gamma^{\rho\sigma} + h^{\rho\sigma})\right]
\frac{1}{\sqrt{-\gamma}}
\frac{\delta L^m}{\delta h^{\rho\sigma}} . \nonumber
\ena
Thus, after using the field equations and removing second derivatives
of $h^{\mu\nu}$, the total energy-momentum
\bea
\theta^{\mu\nu}= -\frac2{\sqrt{-\gamma}} \frac{\delta (L^g + L^m)}
{\delta \gamma_{\mu\nu}} \nonumber
\ena
reduces to 
\bea
\kappa(t^{\mu\nu}|_m + \tau^{\mu\nu}) = \kappa t^{\mu\nu} + 
\left[(\gamma^{\beta\mu}+h^{\beta\nu})(\gamma^{\alpha\nu} +h^{\alpha\nu})-
\frac1{2} (\gamma^{\mu\nu}+h^{\mu\nu})(\gamma^{\alpha\beta}+
h^{\alpha\beta})\right]
\frac{2\kappa}{\sqrt{-\gamma}}\frac{\delta L^m}{\delta h^{\alpha\beta}}.
\nonumber
\ena

Finally, we can write the gravitational field equations in the form similar
to equations (\ref{55}). We take the same linear combination of 
equations (\ref{60}) as was
previously done in (\ref{fe-tem1}). Putting all the terms in the necessary order,
we arrive at the equations equivalent to (\ref{60}):  
\bea
\frac{1}{2}[(\gamma^{\mu\nu}+ h^{\mu\nu})(\gamma^{\alpha\beta}+ 
h^{\alpha\beta})- (\gamma^{\mu\alpha}+h^{\mu\alpha})
(\gamma^{\nu\beta}+h^{\nu\beta})]_{;\alpha;\beta} = 
\kappa (t^{\mu\nu}|_m + \tau^{\mu\nu}). 
\label{68} 
\ena
Thus, in the gravitational field equations, the total energy-momentum
tensor is the source for the generalised d'Alembert operator. Obviously,
the conservation laws $(t^{\mu\nu}|_m + \tau^{\mu\nu})_{;\nu} = 0$ are
satisfied as a consequence of the field equations (\ref{68}).

As a final remark, we should mention that the field-theoretical 
formulation of the general relativity allows also gauge transformations
in addition to arbitrary coordinate transformations. Under gauge
transformations, solutions to the field equations transform into new
solutions of the same equations. In what sense and under which 
conditions the gauge-related solutions are physically equivalent, is 
a deep and nontrivial issue. This question was
partially analyzed in ref. \cite{Gr} but it is outside of the scope of this 
paper. The theory is fully consistent in its mathematical structure
and physical interpretation, if  
the gauge transformations are applied to the gravitational field and
matter variables together (even if we deal only with a couple of 
test particles interacting with the gravitational field and which
are being used in a gedanken experiment).  
We mention the gauge freedom only in order to stress that all
the objects and equations have been derived in arbitrary
gauge, without imposing any gauge conditions.

\subsection{Equivalence with the geometrical Einstein equations}

In the geometrical approach to the general relativity one 
interprets the quantities $g_{\mu\nu}$, introduced
by (\ref{10}), (\ref{g-1}), as the metric tensor of a curved 
space-time (\ref{ds2}). The universal coupling of gravity with matter 
translates into
\bea
L^m=L^m\left[\sqrt{-g} g^{\mu\nu}; (\sqrt{-g} g^{\mu\nu})_{,\alpha}; 
\phi_A; \phi_{A,\alpha}\right]. 
\ena
One can think of this dependence as of manifestation of the Einstein's 
equivalence principle. 

The matter energy-momentum tensor $T_{\mu\nu}$ is now defined
as the variational derivative of $L^m$ with respect to what is now the 
metric tensor: 
$
T_{\mu\nu} = \frac2{\sqrt{-g}} \frac{\delta L^m}{\delta g^{\mu\nu}}
$
The specific form of the $L^m$ allows us to write:
\bea
\frac2{\sqrt{-\gamma}} \frac{\delta L^m}{\delta h^{\mu\nu}} =
\frac{\delta L^m}{\delta (\sqrt{-g}g^{\rho\sigma})} = 
T_{\mu\nu} -\frac1{2}
g_{\mu\nu}g^{\alpha\beta} T_{\alpha\beta}, \label{78}
\ena
Tensor $T^{\mu\nu}$ certainly differs from the tensor $\tau^{\mu\nu}$
defined in the field-theoretical approach, but they are related:
\[
\tau_{\mu\nu} - \frac 1{2}\gamma_{\mu\nu}\gamma^{\alpha\beta}\tau_{\alpha\beta}=
\left(\delta^{\alpha}_{\mu}\delta^{\beta}_{\nu} +
 \frac 1{2}\gamma^{\mu\nu} h^{\alpha\beta} \right)
\left( T_{\alpha\beta}- \frac 1{2}g_{\alpha\beta}g^{\rho\sigma}T_{\rho\sigma}
\right)
\] 

We are now in the position to prove that the field equations 
(\ref{68}) are fully equivalent to the Einstein's geometrical equations.
We know (see (\ref{R-r})) that 
\bea
r_{\mu\nu} = R_{\mu\nu} -\breve R_{\mu\nu}. \label{rR}
\ena
where $\breve R_{\mu\nu}\equiv 0$. 
Combining (\ref{60}) and (\ref{78}) we arrive at the Einstein's geometrical
 equations
\bea
R_{\mu\nu}=\kappa \left( T_{\mu\nu} - \frac1{2}
g_{\mu\nu}g^{\alpha\beta} T_{\alpha\beta} \right). \label{79}
\ena

\section{Conclusions}

We have shown that the field theoretical formulation of the general 
relativity allows us to derive the fully satisfactory gravitational 
energy-momentum tensor $t^{\mu\nu}$  
satisfying all 6 demands listed in the Abstract 
of the paper. Both routes, ``metrical" and ``canonical", lead to one and
the same unique expression (\ref{truemt}). When the gravitational 
field is considered
together with its matter sources, the same strict rules produce the
matter energy-momentum tensor $\tau^{\mu\nu}$ and the modified gravitational
energy-momentum tensor. Both tensors participate at the equal footing in the
nonlinear gravitational field equations (\ref{68}) which are fully 
equivalent to the Einstein's geometrical equations (\ref{79}). These 
strictly defined energy-momentum tensors should be useful in practical 
applications.

\section*{Acknowledgements}

Work of S. V. Babak was partially supported by ORS Award
grant 970047008.

\appendix

\section{Covariant generalisation of the Euler-Lagrange equations}

	The covariant field equations (\ref{cfe}) can be derived in a more 
traditional fashion, when one considers the field variables   
$h^{\mu\nu}$ and their ordinary (not covariant) derivatives 
${h^{\mu\nu}}_{,\tau}$ as functions subject to variation. To emphasize
this fact we rewrite the Lagrangian (\ref{L*}) in the form
\bea
L=L(\gamma^{\mu\nu}, C^{\alpha}_{\ \mu\nu}, h^{\mu\nu}, 
{h^{\mu\nu}}_{,\alpha})~. 
\label{L***}
\ena
Starting from the Lagrangian in this form, one derives the usual field
equations 
\bea
\frac {\partial L}{\partial h^{\mu\nu}} - \left( \frac {\partial L}
{\partial {h^{\mu\nu}}_{,\tau}}\right)_{,\tau} = 0~. \label{afe}
\ena
Since the function $L$ in (\ref{L*}) and (\ref{L***}) is one and the same
function, but written in terms of different arguments, one can relate its
derivatives. The second term in (\ref{afe}) transforms as follows:  
\bea
 \left( \frac {\partial L}
{\partial {h^{\mu\nu}}_{,\tau}}\right)_{,\tau} &=&
 \left( \frac {\partial L}
{\partial {h^{\rho\sigma}}_{;\omega}} \frac {{\partial
{h^{\rho\sigma}}_{;\omega}}}
{{\partial {h^{\mu\nu}}_{,\tau}}} \right)_{,\tau} =
 \left( \frac {\partial L}
{\partial {h^{\mu\nu}}_{;\tau}}\right)_{,\tau} = 
\left( \frac {\partial L}
{\partial {h^{\mu\nu}}_{;\tau}}\right)_{;\tau} +  \nonumber\\
& & \frac {\partial L}{\partial {h^{\sigma\nu}}_{;\tau}}
C^{\sigma}_{\ \mu\tau}+
\frac {\partial L}{\partial {h^{\sigma\mu}}_{;\tau}}
C^{\sigma}_{\ \nu\tau}~.
 \label{d/d,}
\ena
The first term in (\ref{afe}) transforms as 
\bea
 \frac {\partial L}{\partial h^{\mu\nu}}=
\frac {\partial L}{\partial h^{\mu\nu}} +
\frac {\partial L}{\partial {h^{\rho\sigma}}_{;\tau}}
\frac {\partial {h^{\rho\sigma}}_{;\tau}} {\partial h^{\mu\nu}}=
 \frac {\partial L}{\partial h^{\mu\nu}} +
 \frac {\partial L}{\partial {h^{\sigma\nu}}_{;\tau}}
C^{\sigma}_{\ \mu\tau}+
\frac {\partial L}{\partial {h^{\sigma\mu}}_{;\tau}}
C^{\sigma}_{\ \nu\tau}\label{d/d}~.
\ena
Using (\ref{d/d,}) and (\ref{d/d}) in (\ref{afe}) one obtains the required
result (\ref{cfe}). Obviously, the field equations (\ref{afe}) use the
Lagrangian in the form (\ref{L***}), whereas the field equations (\ref{cfe})
use the Lagrangian in the form (\ref{L*}).

\section{Proof of equation (68)}

We need to show in detail that the variational derivative 
of the added term in Lagrangian (\ref{l'}),
calculated at the constraint $\breve R_{\alpha\rho\beta\sigma}=0$, 
result in the last two terms in equation (\ref{tv'}).
Let us introduce a shorter 
notation for the added term stressing its dependence on 
$\gamma_{\mu\nu}$ and derivatives: 
\[ 
\sgm \Lambda^{\alpha\beta\rho\sigma} \breve R_{\alpha\rho\beta\sigma}= 
\Pi(\gamma_{\mu\nu};\gamma_{\mu\nu ,\tau};\gamma_{\mu\nu ,\tau ,\omega}).
\]
Taking into account this dependence, we can write for the variation of
the function $\Pi$: 
\bea
\delta \Pi \equiv 
\left[\frac{\partial \Pi}{\partial \gamma_{\mu\nu}}-
\left(\frac{\partial \Pi}{\partial \gamma_{\mu\nu ,\tau}}\right)_{,\tau}+
\left(\frac{\partial \Pi}{\partial \gamma_{\mu\nu ,\tau,\omega}}
\right)_{,\tau ,\omega}\right] \delta \gamma_{\mu\nu}+
B^{\tau}_{~ ,\tau} \nonumber
\ena
where
\bea
B^{\tau}= \frac{\partial \Pi}{\partial \gamma_{\mu\nu ,\tau}}\delta
\gamma_{\mu\nu}+
 \frac{\partial \Pi}{\partial \gamma_{\mu\nu ,\tau,\omega}}
 \delta \gamma_{\mu\nu ,\omega} -
 \left(\frac{\partial \Pi}{\partial \gamma_{\mu\nu ,\tau,\omega}}
\right)_{ ,\omega}  \delta \gamma_{\mu\nu}. \nonumber
\ena
The coefficient in front of
$\delta \gamma_{\mu\nu}$ defines the variational derivative:  
\bea
\frac{\delta \Pi}{\delta \gamma_{\mu\nu}} =
\frac{\partial \Pi}{\partial \gamma_{\mu\nu}}-
\left(\frac{\partial \Pi}{\partial \gamma_{\mu\nu ,\tau}}\right)_{,\tau}+
\left(\frac{\partial \Pi}{\partial \gamma_{\mu\nu ,\tau,\omega}}
\right)_{,\tau ,\omega}~,  \nonumber
\ena
so that 
\bea
\delta \Pi = \frac{\delta \Pi}{\delta \gamma_{\mu\nu}}  \delta \gamma_{\mu\nu}+
B^{\tau}_{~ ,\tau}~.\nonumber
\ena
In order to find the required variational derivative, we can simply
calculate the variation of $\sgm \Lambda^{\alpha\beta\rho\sigma} 
\breve R_{\alpha\rho\beta\sigma}$ and present it in the form
\bea
\delta (\sgm\Lambda^{\alpha\beta\rho\sigma} 
\breve R_{\alpha\rho\beta\sigma}) = \sgm A^{\mu\nu} \delta\gamma_{\mu\nu} +
(\sgm C^{\tau})_{,\tau}~. \nonumber
\ena
The quantity $\sgm A^{\mu\nu}$ is what we need.  

It is convenient to work with 
$\sgm\Lambda^{\alpha\beta\rho\sigma} \gamma_{\alpha\tau}
\breve R^{\tau}_{\  \rho\beta\sigma}$ instead of 
$\sgm\Lambda^{\alpha\beta\rho\sigma} \breve R_{\alpha\rho\beta\sigma}$.
The variation can be written as  
\bea
\delta (\sgm\Lambda^{\alpha\beta\rho\sigma} \gamma_{\alpha\tau}
\breve R^{\tau}_{\  \rho\beta\sigma}) =
\delta (\sgm\Lambda^{\alpha\beta\rho\sigma} \gamma_{\alpha\tau})
\breve R^{\tau}_{\  \rho\beta\sigma} +
\delta (\breve R^{\tau}_{\  \rho\beta\sigma})
\sgm\Lambda^{\alpha\beta\rho\sigma} \gamma_{\alpha\tau}. \label{p1}
\ena
The first term on the right-hand side of (\ref{p1}) vanishes due to the 
constraint, so we need to focus attention on the second term.
The variation of the Riemann tensor is
\bea
\delta (\breve R^{\tau}_{\  \rho\beta\sigma})=
(\delta C^{\tau}_{\ \rho\sigma})_{;\beta}-
(\delta C^{\tau}_{\ \rho\beta})_{;\sigma}. \label{p2}
\ena
The variation of the Christoffel symbols is 
\bea
\delta C^{\tau}_{\ \rho\sigma} =
\frac 1{2} \gamma^{\tau\lambda} (\delta\gamma_{\lambda\rho ;\sigma} +
\delta\gamma_{\lambda\sigma ;\rho} -\delta\gamma_{\rho\sigma ;\lambda})=
\frac 1{2}(\delta^{\lambda}_{\sigma}\delta^{\alpha}_{\rho}\gamma^{\beta\tau}
+\delta^{\lambda}_{\rho}\delta^{\alpha}_{\sigma}\gamma^{\beta\tau} -
\delta^{\beta}_{\rho}\delta^{\alpha}_{\sigma}\gamma^{\tau\lambda})
{\delta\gamma_{\alpha\beta}}_{;\lambda}.\label{p3}
\ena
One needs to combine (\ref{p3}), (\ref{p2}) and
take into account properties of the symmetry  
of $\Lambda^{\alpha\beta\rho\sigma}$  (see (\ref{symlam})). 
After rearranging the participating terms, one gets 
\bea
 \delta ( \sgm\Lambda^{\alpha\beta\rho\sigma}
\breve R_{\alpha\rho\beta\sigma}) = 
 -\left[ \sgm(\Lambda^{\mu\nu\alpha\beta}+
 \Lambda^{\nu\mu\alpha\beta})_{\alpha;\beta}\right]
\delta \gamma_{\mu\nu} +
\nonumber \\
\left[\sgm(\Lambda^{\alpha\phi\rho\sigma} \gamma_{\alpha\tau}
 - \Lambda^{\alpha\phi\rho\sigma} \gamma_{\alpha\tau})
\delta C^{\tau}_{\ \rho\sigma} +  
\sgm(\Lambda^{\mu\nu\phi\beta}+\Lambda^{\nu\mu\phi\beta})_{;\beta}
\delta\gamma_{\mu\nu}\right]_{;\phi}. \label{pa4}
\ena
The second term on the right-hand side of (\ref{pa4}) is a covariant 
derivative of the vector density, so the covariant derivative can be 
replaced by ordinary derivative, and this term has the form of 
$(\sgm C^{\tau})_{,\tau}$. Thus, we conclude that the sought after 
variational derivative, calculated at the constraint, is  
\bea
\frac{\delta  (\sgm\Lambda^{\alpha\beta\rho\sigma}
\breve R_{\alpha\rho\beta\sigma})}{\delta \gamma_{\mu\nu}} =
- \sgm(\Lambda^{\mu\nu\alpha\beta}+
\Lambda^{\nu\mu\alpha\beta})_{;\alpha ;\beta}~.\nonumber
\ena
Its contribution to the $\kappa\!\stackrel{m}{t}\!^{\mu\nu}|_c$ is 
$
 - (\Lambda^{\mu\nu\alpha\beta} + 
\Lambda^{\nu\mu\alpha\beta})_{;\alpha ;\beta}~
$
what we needed to prove. 

The calculation of (\ref{tv}) can be done in exactly the same way. Namely,
taking the variation of the Lagrangian density (\ref{3}) with respect
to $\gamma_{\mu\nu}$, one obtains

\bea
\delta L^g & =& -\frac{\sgm}{2\kappa} \left\{\left[ \frac 1{2}\gamma^{\mu\nu}
{h^{\rho\sigma}}_{;\alpha}P^{\alpha}_{\ \rho\sigma} + 
[\gamma^{\mu\rho}\gamma^{\nu\sigma} -
\frac1{2}\gamma^{\mu\nu}(\gamma^{\rho\sigma} + h^{\rho\sigma})]
(P^{\alpha}_{\ \rho\beta}P^{\beta}_{\ \sigma\alpha} -
 \frac1{3} P_{\rho}P_{\sigma})\right] \delta \gamma_{\mu\nu}-
 \right.
 \nonumber \\
& & \left.2P^{\alpha}_{\ \rho\sigma}h^{\rho\beta}\delta C^{\sigma}_{\ \alpha\beta}
 \right\}.\label{pa5}
\ena
Using expression (\ref{p3}) and rearranging the participating terms
we arrive at
\bea
\delta L^g &=& -\frac{\sgm}{2\kappa} \left[
\frac1{2} \gamma^{\mu\nu}{h^{\rho\sigma}}_{;\alpha}
P^{\alpha}_{\ \rho\sigma} + [\gamma^{\mu\rho}\gamma^{\nu\sigma} -
\frac1{2}\gamma^{\mu\nu}(\gamma^{\rho\sigma} + h^{\rho\sigma})]
(P^{\alpha}_{\ \rho\beta}P^{\beta}_{\ \sigma\alpha} -
 \frac1{3} P_{\rho}P_{\sigma}) + \right.
\nonumber \\
& & \left. Q^{\mu\nu}\right] \delta \gamma_{\mu\nu}-
\frac{1}{2\kappa}\left[\sgm (P^{\mu}_{ \rho\sigma}h^{\rho\tau}\gamma^{\nu\sigma}
+ P^{\tau}_{ \rho\sigma}h^{\rho\mu}\gamma^{\nu\sigma} -
P^{\nu}_{ \rho\sigma}h^{\rho\mu}\gamma^{\tau\sigma})
\delta \gamma_{\mu\nu}\right]_{,\tau}~.
\ena
Thus, as it is stated in the text (\ref{tv}), the variational derivative is
\bea
-\frac{2 \kappa}{\sgm}\frac{\delta L^g}{\delta \gamma_{\mu\nu}}=
\kappa\!\!\stackrel{m}{t}\!^{\mu\nu} =
\frac1{2} \gamma^{\mu\nu}{h^{\rho\sigma}}_{;\alpha}
P^{\alpha}_{\ \rho\sigma} + [\gamma^{\mu\rho}\gamma^{\nu\sigma} -
 \nonumber \\
 \frac1{2}\gamma^{\mu\nu}(\gamma^{\rho\sigma} + h^{\rho\sigma})]
(P^{\alpha}_{\ \rho\beta}P^{\beta}_{\ \sigma\alpha} -
 \frac1{3} P_{\rho}P_{\sigma}) + Q^{\mu\nu}~.
\ena

\end{document}